\documentclass[nofootinbib,twocolumn,a4paper]{revtex4}
\usepackage{graphicx}
\usepackage[breaklinks=true,colorlinks=true,linkcolor=blue,urlcolor=blue,citecolor=blue]{hyperref}
\usepackage{color}
\usepackage[utf8x]{inputenc}
\usepackage{tikz}
\usepackage{array}
\usepackage{verbatim}
\usepackage{xspace}
\usepackage{amsmath}
\usepackage{orcidlink}

\usepackage{mathrsfs}
\DeclareSymbolFontAlphabet{\mathrsfs}{rsfs}
\newcommand{\scri}{\mathrsfs{I}}
\newcommand{\scripx}{$\scri^+$}
\newcommand{\scrip}{\scripx\xspace}

\newcommand{\aconf}{\bar\Omega}
\newcommand{\papfirst}{Vano-Vinuales:2014koa}
\newcommand{\papgauge}{Vano-Vinuales:2017qij}

\newcommand{\alexthesis}{Vano-Vinuales:2015lhj}

\newcommand{\papbh}{Vano-Vinuales:2023yzs}

\newcommand{\hpt}{h'(\tilde r)}
\newcommand{\rscri}{r_{\!\!\scri}}
\newcommand{\Kc}{K_{CMC}}
\newcommand{\Cc}{C_{CMC}}

\newcommand{\cp}{c}
\newcommand{\const}{\cp^2}

\newcommand{\pr}{\tilde r}
\newcommand{\tort}{\tilde r_*}
\newcommand{\trum}{_{trum}}
\newcommand{\torc}{r_*}
\newcommand{\rtilde}{\tilde r}

\newcommand{\eref}[1]{(\ref{#1})}
\newcommand{\sref}[1]{section~\ref{#1}}
\newcommand{\ssref}[1]{subsection~\ref{#1}}
\newcommand{\aref}[1]{appendix~\ref{#1}}
\newcommand{\fref}[1]{figure~\ref{#1}}

\begin{document}

\title[Conformal diagrams for stationary and dynamical strong-field hyperboloidal slices]{Conformal diagrams for stationary and dynamical strong-field hyperboloidal slices}

\author{Alex Vañó-Viñuales\,\orcidlink{0000-0002-8589-006X}}
\address{Centro de Astrof\'{\i}sica e Gravita\c c\~ao - CENTRA, Departamento de F\'isica, Instituto Superior T\'ecnico IST, Universidade de Lisboa UL, Avenida Rovisco Pais 1, 1049-001 Lisboa, Portugal}
\email{alex.vano.vinuales@tecnico.ulisboa.pt}

\begin{abstract}
Conformal Carter-Penrose diagrams are used for the visualization of hyperboloidal slices, which are smooth spacelike slices reaching null infinity. The focus is on the Schwarzschild black hole geometry in spherical symmetry, whose Penrose diagrams are introduced in a pedagogical way. The stationary regime involves time-independent slices. In this case, different options are given for integrating the height function -- the main ingredient for constructing hyperboloidal foliations. The dynamical regime considers slices changing in time, which are evolved together with the spacetime using the eikonal equation. It includes the relaxation of hyperboloidal Schwarzschild trumpet slices and the collapse of a massless scalar field into a black hole, for which Penrose diagrams are presented. 
\end{abstract}

\maketitle

\section{Introduction}

Conformal Carter-Penrose diagrams\footnote{Also called conformal diagrams, Carter-Penrose diagrams or, simply, Penrose diagrams. All these names will be used as synonyms throughout the text.} are a very useful tool to visualize the causal structure of spacetimes, as they allow to depict infinitely large distances in space and time in a finite domain. They employ Penrose's conformal compactification \cite{PhysRevLett.10.66} to compress along the null directions and provide a representation of the spacetime where 
its causal structure remains the same. 
For some references regarding the construction of Carter-Penrose diagrammatic representations of spherically symmetric spacetimes (and of Kerr along its axis of symmetry \cite{PhysRev.141.1242}) see for instance \cite{10.1063/1.1665393}, \cite{Hawking:1973uf}, appendix C in \cite{Dafermos:2003yw}, \cite{Griffiths:2009dfa} and chapter 6 in \cite{Kroon:2016ink}. 
The comprehensive treatment in \cite{Schindler:2018wbx} focuses on strongly spherically symmetric spacetimes (those with the form of \eref{ed:} and \eref{ein:lielphys}) and spacetimes with (evolving) nontrivial matter distributions; its global tortoise function is followed in \cite{Cederbaum:2023keu}, where their ordinary-differential-equation approach is applied to photon surfaces. 
This work will restrict itself to depictions of spherically symmetric spacetimes. However, the projection diagrams developed in \cite{Chrusciel:2012gz} are suitable for geometries with less symmetries, and applied to create conformal diagrams of e.g.~the Kerr(-Newman) spacetime with and without cosmological constant.
See also the work in \cite{Gibbons:1999uv} for a rotating extreme black hole in five spacetime dimensions.

Besides their clarity and aesthetic elegance, Penrose diagrams have also played an important role in clarifying properties of spacetimes. 
For instance, in \cite{10.1063/1.524400} they were used to depict numerically-constructed hypersurfaces of constant mean curvature in Schwarzschild. Similarly in \cite{Beig:2005ef} for a cosmological model derived from the Kottler-Schwarzschild-deSitter spacetime. A systematic study of hyperboloidal foliations (see below for a brief description) in conformal diagrams is included in \cite{Zenginoglu:2007it}. In \cite{Ohme:2009gn} they were used to depict (figure 11) the stationary hyperboloidal slices  of the Schwarzschild spacetime found, while in \cite{Dennison:2014sma} (figure 2) the same was done for a family of non-hyperboloidal family of trumpet solutions. 
As tools for the visualization of numerically evolved spacetimes, conformal diagrams were key in understanding the numerical evolution from wormhole initial data to effectively trumpet slices with the moving puncture gauge in \cite{Hannam:2008sg}. 
Evolution of spacetime slices of a collapsing star were depicted in figure 7 of \cite{Thierfelder:2010dv} as a (non-compactified) causal spacetime diagram, and they were determined using the eikonal equation. 
Figure 10 in \cite{Khirnov:2018rjc} shows conformal diagrams for sub-critical and super-critical collapse scenarios of Brill wave initial data. 
Penrose diagrams were also used to give a complete impression of the hyperboloidal slicings considered/determined in \cite{\papbh}. 
Literature on mathematical relativity may also aid their explanations with Penrose diagrams, such as e.g.~\cite{Ortiz:2011jw,Garcia-ParradoGomez-Lobo:2017wmd,Minucci:2023xcw}. Further recent work that makes use of conformal diagrams is \cite{Joshi:2023ugm,Bozzola:2023daz}.
This work will focus on the use of Penrose diagrams for the visualization of a specific type of slices used for the numerical evolution of spherically symmetric spacetimes including a non-charged black hole. 

The hyperboloidal initial value problem \cite{friedrich1983,lrr-2004-1,Friedrich:2003fq} consists of evolving the Einstein equations on hyperboloidal slices \cite{Zenginoglu:2007jw} -- spacelike slices that reach null infinity. Future null infinity (\scrip) corresponds to the collection of end points of future-directed null geodesics (see its location in the Penrose diagrams included in this work, such as figures \ref{basicmin} and \ref{basicschw}).
It is there that global properties of spacetimes are defined and it also corresponds to the correct idealisation \cite{Barack:1998bv,Leaver1986,PhysRevD.34.384} of observers of astrophysical events, such as the gravitational wave interferometers. The great advantages of the hyperboloidal approach are that radiation travelling at the speed of light can be unambiguously extracted directly at \scrip from numerical simulations, that the treatment of the outer boundary, a difficult issue in numerical simulations, simplifies enormously, and that the spacelike character of hyperboloidal slices provides more flexibility than the characteristic approach. Its relative shortcoming is that the hyperboloidal initial value problem provides only a semi-global solution and does not give any information about spacelike infinity ($i^0$). 

This work was motivated by studies on the behaviour of hyperboloidal slices involved in strong-field spherically symmetric experiments by the author. For more information about the setup, which follows the conformal compactification approach \cite{PhysRevLett.10.66}, see \cite{\papfirst,\papgauge,\papbh,\alexthesis} and references therein. 
The focus here is on the numerical construction of the hyperboloidal slices for the diagrams, as well as their evolution together with the spacetime. New results are the visualization of hyperboloidal Schwarzschild trumpet slices relaxing (\ssref{trumpevol}) to their stationary state (\ssref{statio}) and of hyperboloidal slices during the collapse of a massless neutral scalar field (\ssref{colevol}). To make the construction of the diagrams as straightforward as possible, this paper is structured in a pedagogical manner. It can also be used as a practical introduction to building conformal diagrams of the Schwarzschild spacetime. Part of the prescriptions can be extended to Reissner-Nordstr\"om (with or without cosmological constant). 

This paper is organised as follows. The basic coordinate transformations used to construct a conformal diagram of a spherically symmetric asymptotically flat spacetime are included in \sref{basic}. Section \ref{sta} introduces the height function approach for the description of stationary spacetime foliations and explains several options for its calculation in the hyperboloidal case (one of them being located in \aref{apextort}). The dynamical regime is considered in \sref{dyn}, describing how to use the eikonal equation to evolve the profiles of the spacetime slices and showing results for relaxing trumpet slices and the collapse of a scalar field to a black hole. Some concluding remarks are gathered in \sref{conclu}. Appendix \ref{apix} indicates how to transform between the physical and conformally rescaled quantities used in this work.

The metric signature is $(-,+,+,+)$ and the fundamental constants are set to $G=c=1$. The convention for the sign of the extrinsic curvature is that of Misner, Thorne, and Wheeler \cite{Misner1973}: a negative value means expansion of the normals. This is why the constant parameter $\Kc$ introduced in \eref{hdef} is negative for hyperboloidal slices reaching future null infinity.

Supplementary material \cite{penroserepo} accompanying this work includes a Mathematica notebook that constructs the Penrose diagrams as described and included here, as well as numerical data to create the plots for the dynamical regime and animations for those dynamical spacetimes. Other useful resources are the creation of Carter-Penrose diagrams with \texttt{SageMath} \cite{sage}, \texttt{python} \cite{penpyth}, and with the \LaTeX package \texttt{Tikz} and \texttt{python} \cite{pentikz}. 

\section{Basic construction}\label{basic}

\subsection{Conformal compactification}

The coordinate transformations used to build a Carter-Penrose diagram follow a compactification along the null directions. A conformal rescaling \eref{conf} of the resulting metric is useful for some applications, but it is not necessary for the conformal diagrams. For a step-by-step derivation follow e.g.~section 11.1 in \cite{Wald}. Here only the basic ingredients are provided. Starting from some functions $\tilde T(\tilde t, \rtilde)$ and $\tilde R(\tilde t, \rtilde)$ of the original\footnote{Original in the sense that they correspond to the time and radius in the chosen line element, and that they are not compactified.} time $\tilde t\in(-\infty,\infty)$ and radial $\rtilde\in[0,\infty)$ coordinates, perform the transformations
\begin{subequations}\label{basictrafo}
\begin{eqnarray}
 \tilde U = \tilde T - \tilde R, \ && \tilde V = \tilde T+\tilde R, \label{basic1} \\
 U=\arctan\tilde U, \ && V=\arctan\tilde V, \label{basic2} \\
 T = \frac{V+U}{2}, \ && R = \frac{V-U}{2}, \label{basic3}
\end{eqnarray}
\end{subequations}
where all quantities are functions of the time $\tilde t$ and radius $\rtilde$. 
The $\arctan$ function above is the usual choice for the compactification, but other options are also possible. 
In the case of Minkowski spacetime (with metric $d\tilde s^2=-d\tilde t^2+d\rtilde^2$ suppressing the angular part), the choice is simply $\tilde T=\tilde t$ and $\tilde R = \rtilde$. As described in the next subsection, black hole spacetimes use a different choice of $\tilde T$ and $\tilde R$. 
To display the Penrose diagram, $R$ is plotted on the horizontal axis and $T$ on the vertical one, within their ranges of $[-\pi/2,\pi/2]$. From the original infinite ranges of $\tilde t$ and $\rtilde$, we have obtained a finite extent for $T$ and $R$, so bringing infinity to a finite location. 

\begin{figure}
\center
\includegraphics[width=0.48\linewidth]{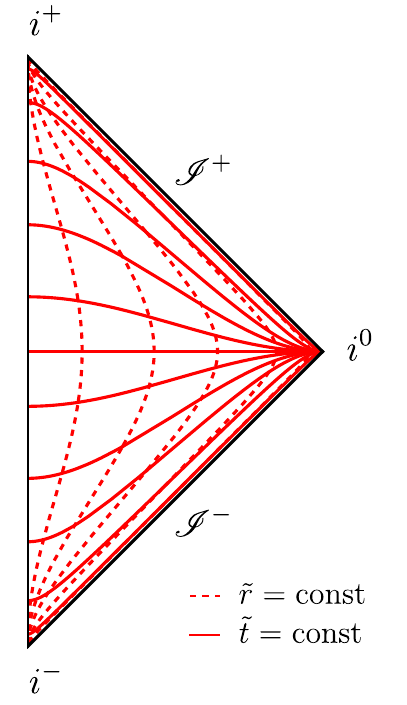}
\includegraphics[width=0.48\linewidth]{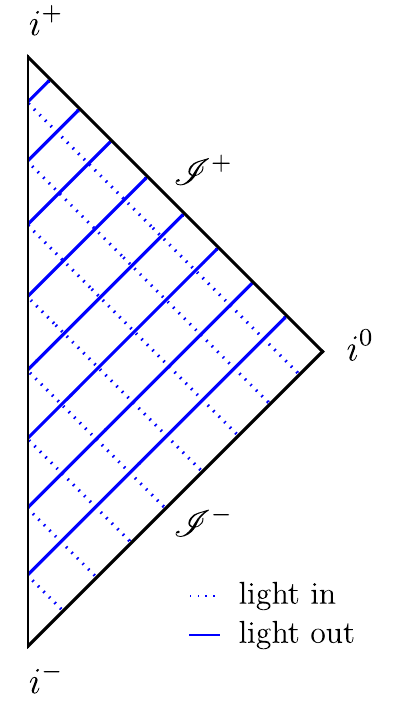}
\caption{Carter-Penrose diagrams of Minkowski spacetime: on the left, slices with constant values of $\tilde t$ and $\rtilde$ coordinates are depicted, while on the right, constant values of the null coordinates $\tilde U\equiv \tilde u = \tilde t-\rtilde$ and $\tilde V\equiv \tilde v = \tilde t+\rtilde$ determine the foliations. The angular coordinates do not play a role in the conformal compactification and have been left unchanged. Thus, each point in a Penrose diagram corresponds to a 2-sphere whose radius is the value of $\tilde r$ at that point.}\label{basicmin}
\end{figure}
Plotting $T$vs.$R$ for Minkowski serves to illustrate the basic asymptotic components of an asymptotically flat spacetime. On the left in \fref{basicmin}, spacelike slices at constant time $\tilde t$ correspond to the solid red lines. They extend between the vertical line on the left (the axis of symmetry, corresponding to $\rtilde=0$) and spacelike infinity $i^0$ (for $\rtilde\to\infty$). Timelike slices at constant radius $\rtilde$ are denoted by the red dashed lines: they all start at past timelike infinity $i^-$ for $\tilde t\to-\infty$ and end at future timelike infinity $i^+$ for $\tilde t\to\infty$. On the right diagram, (radially-directed, flat) null geodesics, the paths followed by light rays, are shown. Ingoing null slices, in blue dotted pattern, correspond to constant values of the advanced null coordinate $\tilde v = \tilde t+\rtilde$, and they enter the domain through past null infinity ($\scri^-$). Slices with constant retarded null coordinate $\tilde u = \tilde t-\rtilde$ represent outgoing light rays that leave the spacetime through \scrip. 

\subsection{Kruskal-Szekeres-like coordinates for black hole spacetimes} 

One way to deal with the coordinate singularity at the horizon in the case of a black hole spacetime is to introduce the tortoise radial coordinate $\tort$. This allows to write the metric in a manifestly conformally flat form in 2+1+1. Consider the following general ansatz for a static metric in spherical symmetry (satisfying the Einstein equations and the required conditions for the stress-energy tensor \cite{Salgado:2003ub}) with the angular part omitted,
\begin{equation}\label{ed:}
d\tilde s^2 = -A(\tilde r)d\tilde t^2+\frac{1}{A(\tilde r)}d\tilde r^2 \equiv A(\tilde r)\left(-d\tilde t^2+d\tort^2\right) , 
\end{equation}
where the tortoise coordinate has been introduced as
\begin{equation}\label{dtortdr}
d\tort = \frac{d\rtilde}{A(\rtilde)} . 
\end{equation}
These coordinate changes are again textbook material. For a slightly more detailed derivation see e.g.~appendix A in \cite{\alexthesis}. Further use of the tortoise coordinate in calculations for this work is collected in \aref{apextort}.  
Similarly as before, the main transformations happen along the null directions \eref{ksl2}, 
\begin{subequations}\label{ksl}
\begin{eqnarray}
\tilde u=\tilde t-\tort , \ && \tilde v=\tilde t+\tort ,\label{ksl1}\\
 u=-e^{-\frac{\tilde u}{B}} , \ && v=e^{\frac{\tilde v}{B}}  , \label{ksl2} \\
 \tilde T=\frac{v+u}{2} , \ && \tilde R=\frac{v-u}{2}, \label{ksl3}
\end{eqnarray}
\end{subequations}
with $B$ a given function of the parameters specific to $A(\rtilde)$. The choice $B=4M$ in the Schwarzschild spacetime gives the Kruskal-Szekeres coordinates\footnote{The value of $B$ is fixed by requiring that the metric be devoid of coordinate singularities at $\tilde r=2M$. Given that the $\tilde t-\tilde r$ (or $\tilde u-\tilde v$) part of the metric is conformally flat, its information is encoded in the $\omega$ conformal factor in $d\tilde s^2 = \omega^2 du dv$, with $B$ chosen so that $\omega$ is finite and continuous over the horizon.}, see \eref{TRSchw}. Note the similar structure of \eref{ksl} with respect to \eref{basictrafo}. 

In terms of the original $\tilde t$ and $\tilde r$, \eref{ksl}'s final result is 
\begin{equation}\label{ed:TR}
\tilde T=e^{\frac{\tort}{B}}\sinh\left(\frac{\tilde t }{B}\right) , \ \tilde R=e^{\frac{\tort}{B}}\cosh\left(\frac{\tilde t }{B}\right) , 
\end{equation}
where $\tort$ can be expressed in terms of $\rtilde$ from \eref{dtortdr}. 
In the case of the Schwarzschild spacetime, the tortoise coordinate introduced in \eref{dtortdr} integrates to 
\begin{subequations}\label{ptortexpr}
\begin{eqnarray}
&\pr>2M: \qquad \tort = \pr + 2M \log\left(\frac{\pr}{2M}-1\right), \\
&\pr<2M: \qquad \tort = \pr + 2M \log\left(1-\frac{\pr}{2M}\right),
\end{eqnarray}
\end{subequations}
and \eref{ed:TR} takes the explicit form in Kruskal-Szekeres coordinates
\begin{subequations}\label{TRSchw}
\begin{equation}\label{ed:TRSchwout}
\tilde r>2M:
\left\{\begin{array}{l}
\tilde T=\sqrt{\frac{\tilde r}{2M}-1}\ e^{\frac{\tilde r}{4M}}\sinh\left(\frac{\tilde t }{4M}\right) , \\
\tilde R=\sqrt{\frac{\tilde r}{2M}-1}\ e^{\frac{\tilde r}{4M}}\cosh\left(\frac{\tilde t }{4M}\right) , 
\end{array}\right.
\end{equation}
\begin{equation}\label{ed:TRSchwin}
\tilde r<2M:
\left\{\begin{array}{l}
\tilde T=\sqrt{1-\frac{\tilde r}{2M}}\ e^{\frac{\tilde r}{4M}}\cosh\left(\frac{\tilde t }{4M}\right) , \\
\tilde R=\sqrt{1-\frac{\tilde r}{2M}}\ e^{\frac{\tilde r}{4M}}\sinh\left(\frac{\tilde t }{4M}\right) . 
\end{array}\right.
\end{equation}
\end{subequations}
Conformal diagrams are built by blocks, in the form of diamonds or triangles, whose boundaries (or ``seams'', as called in \cite{10.1063/1.1665393}) are determined by the locations where $A(\rtilde)$ changes sign. This also implies a change in the character of the coordinates, with $\tilde t$ becoming spacelike and $\rtilde$ becoming timelike when $A(\rtilde)<0$. There are also considerations to take into account in how the blocks can be glued together, see for instance \cite{10.1063/1.1665393,Chrusciel:2012gz}. Here only the outer spacetime \eref{ed:TRSchwout} and the interior of the black hole horizon \eref{ed:TRSchwin} will be considered, so no further blocks are needed. For more details see e.g.~appendix A in \cite{\alexthesis}, where the complete derivation for Schwarzschild and Reissner-Nordstr\"om is included. 

\begin{figure}
\center
\includegraphics[width=0.98\linewidth]{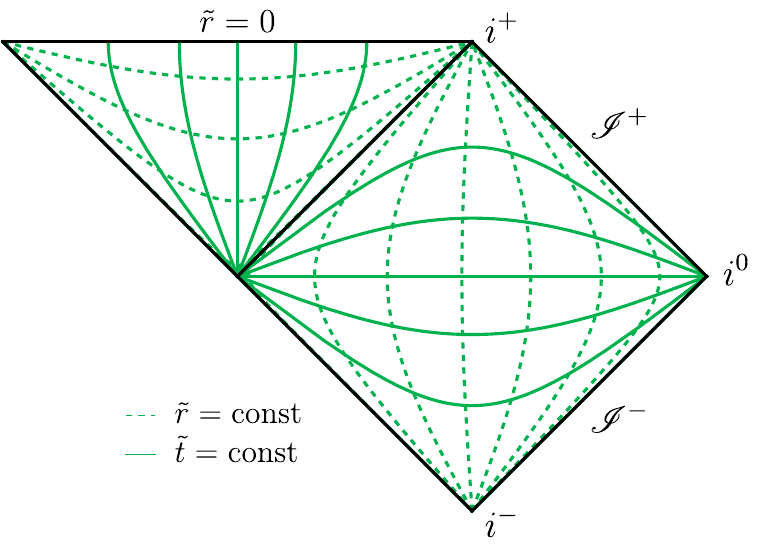}
\caption{Carter-Penrose diagram showing the Schwarzschild spacetime: slices with constant values of $\tilde t$ and $\rtilde$ coordinates are depicted. The singularity is located at $R=0$, at the top of the diagram, while the horizon $R=2M$ corresponds to the black diagonal line separating the two blocks: the inside of the horizon is the one in the triangular shape, while the outer spacetime is the diamond.}\label{basicschw}
\end{figure}

To finish building the Penrose diagram of the Schwarzschild black hole, $\tilde T$ and $\tilde R$ above are to be substituted into \eref{basic1}. The result is shown on \fref{basicschw}. 

\subsection{Change in the time coordinate}

The Einstein equations are commonly expressed as an initial value formulation. For this, spacetime is sliced along hypersurfaces with constant value of a given parameter, usually identified with the time coordinate. 
One simple way to modify the slicing is to change the time parameter whose level sets provide the spacetime slices. The original time coordinate $\tilde t$ is transformed to a new time $t$ via
\begin{equation}\label{ttrafo}
\tilde t = t + h
\end{equation}
where $h$ is the height function \cite{10.1063/1.1666384,Malec:2003dq,Calabrese:2005rs}. 
It depends only on spatial coordinates, so as to preserve the timelike Killing field. 
In the present setup, $h$ is a function of only the radial coordinate. 
The effect of the height function is to introduce a mixing between the time and radial coordinates, so that the level sets of the new $t$ correspond to hypersurfaces that better suit the problem at hand. A clear example are the maximal trumpet slices in \cite{PhysRevD.7.2814}, considered in further works such as \cite{Baumgarte:2007ht,Hannam:2008sg}, where the constant $t$ hypersurfaces end inside the black hole's horizon, but avoid the singularity. 
In the context of hyperboloidal foliations, the height function is mainly used to modify the asymptotic behaviour of the time coordinate. 
In this case, the height function must satisfy $dh/d\tilde r < 1$ everywhere except asymptotically, where $dh/d\tilde r|_\scri=1$ holds. This is equivalent to the hyperboloidal time becoming the retarded time $u$ at infinity. Hyperboloidal slices are thus spacelike, but extend to \scrip. This can be seen for instance in the hyperboloidal slices depicted in \fref{minkcmcpert}. 
Depending on the choice of $h$, the considered spacetime will be sliced along different $t=$ constant hypersurfaces. The next section describes the numerical calculation of $h$ for stationary hyperboloidal Schwarzschild trumpet slices. 

\section{Stationary spacetimes}\label{sta}

This applies to spacetimes with a timelike Killing vector, or equivalently those spacetimes whose metric components can be written as explicitly time-independent functions. 
A stationary spacetime that is spherically symmetric, as those under consideration here, is also static. 
Given that the numerical experiments that motivated this work are carried out on compactified spatial coordinates, the construction of Penrose diagrams will be described for both the physical and the compactified domains. 
The height function approach is employed to determine the spacetime slices for the Carter-Penrose diagrams that correspond to the metric initial data used in the numerical evolutions. It is also required to create diagram initial data for the dynamical spacetimes in \sref{dyn}. 
The spacetime under consideration will be Schwarzschild, for which closed-form expressions for the height function exist for a specific choice of slicing. These will be used when illustrating derivations on the physical radial coordinate, while either compactified closed-form expressions or numerical data will be employed when calculating in the compactified domain. 

\subsection{Height function in physical domain}\label{hphys}

Consider the following ansatz for the initial metric, which is general enough to consider flat spacetime, the Schwarzschild and Reissner-Nordström geometries, and the addition of a cosmological constant. 
Under the time change \eref{ttrafo} with $h=h(\rtilde)$ and $ds\sigma^2\equiv d\theta^2+\sin^2\theta d\phi^2$, the line element transforms as
\begin{subequations}\label{ein:lielphysboth}
\begin{eqnarray}\label{ein:lielphys}
d\tilde s^2 &=& -A(\tilde r)d\tilde t^2+\frac{1}{A(\tilde r)}d\tilde r^2+\tilde r^2 d\sigma^2 \\ 
&=&  -A\,dt^2 -2A\,\partial_{\tilde r}h\,dt\,d\tilde r+\frac{1-(A\,\partial_{\tilde r}h)^2}{A\,}d\tilde r^2 \nonumber \\ && +\, \tilde r^2 d\sigma^2. \label{ein:lielphysh}
\end{eqnarray}
\end{subequations}

The height function's purpose is to make the constant $t$ slices become null asymptotically. One simple way to achieve this is to consider constant-mean-curvature (CMC) slices, where the trace of the physical extrinsic curvature is set to a constant non-zero value. The procedure (see~\cite{10.1063/1.524400,Gentle:2000aq,Malec:2003dq} for more details) involves expressing the unit normal $\tilde n^a$ to the hypersurface in terms of the metric \eref{ein:lielphysh} and use it to calculate the expression for the trace of the physical extrinsic curvature, which is set here to the (negative) parameter $\Kc$. 
\begin{eqnarray}\label{hdef}
\Kc &=& -\frac{1}{\sqrt{-\tilde g}}\partial_a\left(\sqrt{-\tilde g}\,\tilde n^a\right) \nonumber \\ &=& -\frac{1}{r^2}\partial_r\left[ r^2 \frac{A^{3/2}(\tilde r)\,\hpt}{\sqrt{1-\left(A(\tilde r)\hpt\right)^2}} \right] .
\end{eqnarray}
Introducing $\Cc$ as an integration constant, the first derivative of the height function is isolated to give 
\begin{equation}\label{hfunc}
\hpt = - \frac{\frac{\Kc\,\tilde  r}{3} + \frac{\Cc}{\tilde r^2} }{A(\tilde r)\sqrt{A(\tilde r)+\left(\frac{\Kc\,\tilde  r}{3} + \frac{\Cc}{\tilde r^2}\right) ^2}}.
\end{equation}
When considering numerical initial data in \ssref{hcomp}, it will be useful to determine the height function from a given metric (assumed to be a function of a suitable hyperboloidal time coordinate). Let us thus introduce that infrastructure here. The generic metric 
\begin{equation}
d\tilde s^2 = \tilde g_{\mu\nu}\,dx^\mu\,dx^\nu =  \tilde g_{tt} dt^2 + 2 \tilde g_{ti} dt\,dx^i + \tilde g_{ij} dx^i\,dx^j 
\end{equation} 
is written in spherical symmetry as
\begin{equation}\label{gphys}
d\tilde s^2 = \tilde g_{tt} dt^2 + 2 \tilde g_{t\rtilde}\,dt\,d\rtilde +  \tilde g_{\rtilde\rtilde}\, d\rtilde^2 +  \tilde g_{\theta\theta}\, \rtilde^2\, d\sigma^2 .
\end{equation}
Note that the $\rtilde^2$ part has not been included in the definition of $\tilde g_{\theta\theta}$. This is merely for convenience. 
Comparing the mixed $dt\,d\rtilde$ term in \eref{ein:lielphysh} and in \eref{gphys}, the relation for the first derivative of the height function is (with $g_{tt}=-A$)
\begin{equation}\label{hpinteg}
h'(\rtilde)\equiv\partial_{\rtilde} h(\rtilde)=\frac{\tilde g_{t\rtilde}}{\tilde g_{tt}}. 
\end{equation}

In spherical symmetry, $A(\rtilde)$ can describe a black hole with mass parameter $M$ and charge $Q$, and include a cosmological constant $\Lambda$
\begin{equation}
A(\tilde r)=1-\frac{2M}{\rtilde}+\frac{Q^2}{\rtilde^2}-\frac{\Lambda}{3}\rtilde^2 . 
\end{equation}
From now onward we will restrict ourselves to the Schwarzschild geometry ($Q=0$ and $\Lambda=0$), and use $M=1$ and $\Kc=-1$ when setting numerical values for the parameters. The behaviour of the hyperboloidal slices inside of the horizon depends on the value of $\Cc$. Of special relevance is a critical value of $\Cc$ ($\approx 3.11$ for the chosen values of $M$ and $\Kc$), which provides slices that are continuous everywhere and avoid the singularity by adopting a trumpet geometry \cite{Hannam:2006vv,PhysRevD.7.2814,Baumgarte:2007ht}. The critical value of $\Cc$ providing hyperboloidal CMC trumpet slices will be the choice here due to its suitability for numerical experiments. To see the behaviour of slices with non-critical values of $\Cc$, see e.g.~figure 3 in \cite{\papbh} (and figures 3.11-14 in \cite{\alexthesis} for the Reissner-Nordstr\"om spacetime). 

The companion Mathematica notebook in \cite{penroserepo} constructs diagrams in the physical domain using the closed-form expression \eref{hfunc}, in order to illustrate how the procedure works with the uncompactified radial coordinate. There are no obvious advantages in transforming numerical output from the compactified domain to the physical one to build the conformal diagrams, but it may be required in some applications. Appendix \ref{apix} describes how to transform conformally compactified numerical data to the physical domain.

\subsubsection{Integrating $h$ in physical radial coordinate $\rtilde$}\label{hcalcp}

For simplicity, let us consider the construction for the Minkowski spacetime first. The height function for a CMC slice is obtained by setting $A(\tilde r)=1$ and $\Cc=0$~\footnote{If $\Cc\neq 0$, the slices become null at the origin. They asymptote to $\tilde U\equiv \tilde u = \tilde t-\rtilde =$ const there (outgoing null direction) for $\Cc<0$ and $\tilde V\equiv \tilde v = \tilde t+\rtilde =$ const (ingoing null direction) for $\Cc>0$.} in \eref{hfunc}, and integrates to the closed-form expression
\begin{equation}\label{hminkp}
h(\tilde r)=\sqrt{\left(\frac{3}{\Kc}\right)^2+{\tilde r^2}} + \frac{3}{\Kc}. 
\end{equation} 
The last term is an integration constant added to make the above expression zero at the origin $\rtilde=0$. 
The Penrose diagram is constructed by substituting $\tilde R = \rtilde$ and $\tilde T=\tilde t = t + h(\rtilde)$ with the above height function into \eref{basic1}. Constant values of $t$ give the hyperboloidal slices. The result will look like the purple slices in \fref{minkcmcpert}.

\begin{figure}
\center
\includegraphics[width=0.96\linewidth]{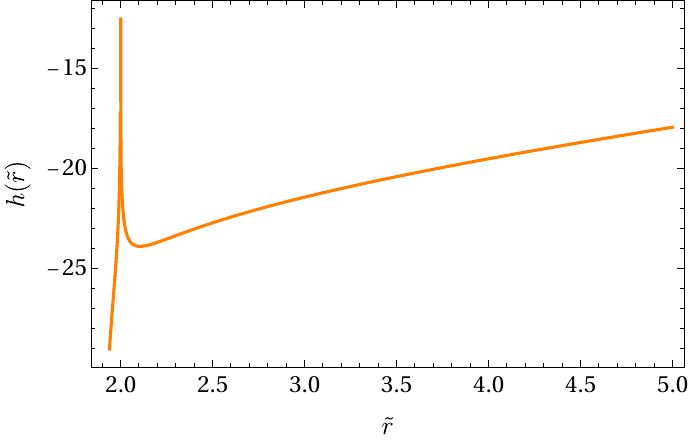}
\caption{Hyperboloidal height function expressed in terms of the physical (in the sense of uncompactified) radial coordinate. It diverges at the horizon $\rtilde=2M$ if expressed in terms of the physical Schwarzschild radial coordinate $\rtilde$. Asymptotically it grows linearly with $\rtilde$, the same behaviour of the Minkowski CMC height function \eref{hminkp}. Compare to the solid line in figure 4 in \cite{\papbh}.}\label{heightp}
\end{figure}

In the case of the Schwarzschild geometry, expressions \eref{TRSchw} are to be substituted into \eref{basictrafo}. Here, \eref{hfunc} is to be integrated numerically so it can be used in the time change \eref{ttrafo}. The zero of $A(\rtilde)$ at the horizon $\rtilde=2M$ in the currently used Schwarzschild coordinates induces a divergence in the derivative of the height function \eref{hfunc}. Hence, the integration has to be performed separately in the domains $\pr\trum<\rtilde<2M$ and $2M<\rtilde<\infty$, with the condition that $h|_{\tilde r = 2M-\epsilon}=h|_{\tilde r = 2M+\epsilon}$ for $\epsilon$ as small as numerically possible. The value of $\pr\trum$ ($\approx 1.91$ for $M=1$, $\Kc=-1$ and $\Cc\approx 3.11$) corresponds to the location of the trumpet's throat and it is the closest the foliation gets to the singularity in terms of the Schwarzschild radial coordinate. For more information about $\pr\trum$, see e.g.~\cite{Baumgarte:2007ht} for the maximal case and subsection IV D in \cite{\papbh} for the hyperboloidal one, where $\pr\trum$ is denoted as $R_0$. The integrated height function is depicted in \fref{heightp}. Its integration constant has been set so that the height function coincides with \eref{hminkp} for $\rtilde\to\infty$. The same criterion for setting the integration constants will be used in the rest of this section.

The conformal diagram corresponding to \fref{heightp}'s height function is included in \fref{penphysboth}, including the location of $\pr\trum$ in a thick dashed orange line. The hyperboloidal slices constructed are depicted as orange solid lines; it can be clearly seen that the lower slices are not smooth around the horizon, due to errors in the calculation of the height function there. The inaccuracies included by the coordinate singularity at the horizon motivate a more suitable choice of coordinates. Using the tortoise coordinate as in \aref{apextort} is a considerable improvement, but the following method deals with the singularity better.

\subsubsection{Integrating $\Delta h$ in physical radial coordinate $\rtilde$}\label{hcalcpdh}

\begin{figure}
\center
\includegraphics[width=0.96\linewidth]{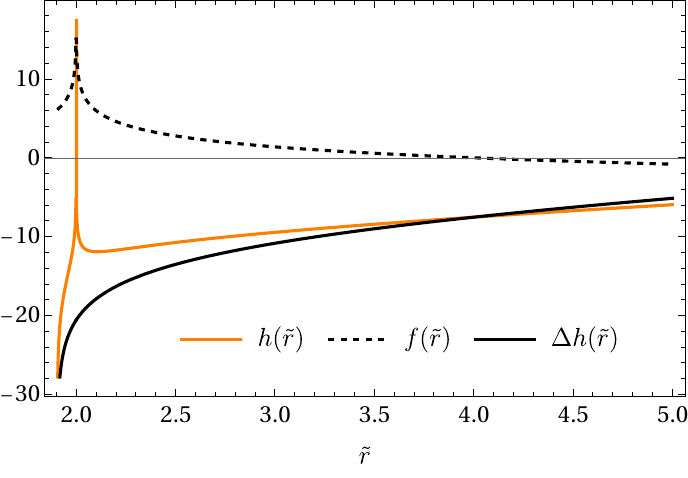}
\caption{Height functions $h$ (orange) and $f$ (dashed black), as well as their difference $\Delta h$ (solid black), expressed in terms of the physical radial coordinate. While $h(\rtilde)$ and $f(\rtilde)$ diverge at the horizon $\rtilde=2M$, $\Delta h(\rtilde)$ is finite and smooth there. The integration constant has been set so that $h(\rtilde)$ behaves asymptotically like the Minkowski height function \eref{hminkp}.}\label{heightpdh}
\end{figure}

The metric in Kerr-Schild form \cite{10.1063/1.1664769} is given by
\begin{equation}
\tilde g_{ab} = \tilde \eta_{ab} + \tilde C \tilde l_a \tilde l_b, 
\end{equation}
where $\tilde \eta_{ab}$ is the flat metric, $\tilde l_a$ is a null vector and for Schwarzschild $\tilde C=2M/\rtilde$. An advantage of this way of expressing the metric is that it is devoid of coordinate singularities at the horizon. 
One way to obtain the Schwarzschild metric in Kerr-Schild coordinates is to perform the time transformation
\begin{equation}\label{tftrafo}
\tilde t = \hat t + f
\end{equation}
with a height function given by (see \cite{schwks} or (5a) in \cite{Kerr:2023rpn})
\begin{subequations}\label{fksexpr}
\begin{eqnarray}
&\pr>2M: \qquad f(\rtilde) = -2M \log\left(\frac{\pr}{2M}-1\right), \\
&\pr<2M: \qquad f(\rtilde) = -2M \log\left(1-\frac{\pr}{2M}\right). 
\end{eqnarray}
\end{subequations}
Here the sign has been chosen so that the regions of interest of the Schwarzschild spacetime (those shown in \fref{basicschw}) are covered. 
The height function $f(\rtilde)$ encodes the coordinate divergence at the horizon and is depicted with the black dashed line in \fref{heightpdh}. 
The idea is to separate the height function $h$ in \eref{ttrafo} into a part holding the divergence ($f$) and a part regular at the horizon ($\Delta h$) as
\begin{equation}
h = \Delta h + f. 
\end{equation}
The time change \eref{ttrafo} using \eref{tftrafo} is thus rewritten as 
\begin{equation}
\tilde t = t + h = t+\Delta h + f, \quad \textrm{with} \quad \hat t = t + \Delta h.
\end{equation}
Applying a derivative to $f(\rtilde)$'s expression \eref{fksexpr} yields $f'(\rtilde) = (1-\frac{\rtilde}{2M})^{-1}$. After some algebra, this expression is written in terms of $A$ (from \eref{ein:lielphysh}) and the metric components as
\begin{equation}\label{fprphysexpr}
f'(\rtilde) = -\frac{1-A}{A} = \frac{1+\tilde g_{tt}}{\tilde g_{tt}}. 
\end{equation}
Finally, putting \eref{hpinteg} and \eref{fprphysexpr} together provides the expression for $\Delta h$ that is to be integrated numerically: 
\begin{equation}\label{dhpinteg}
\Delta h'(\rtilde)\equiv\partial_{\rtilde}\Delta h(\rtilde) = h'(\rtilde) -f'(\rtilde)=\frac{\tilde g_{t\rtilde}-\tilde g_{tt}-1}{\tilde g_{tt}}. 
\end{equation}
The result is shown on \fref{heightpdh}. The black solid line shows the integrated $\Delta h$: it is smooth throughout the integration domain and only diverges at the location of the trumpet (at the left boundary), same as $h$. This makes the numerical integration much simpler and reliable. 

\begin{figure}
\center
\includegraphics[width=0.96\linewidth]{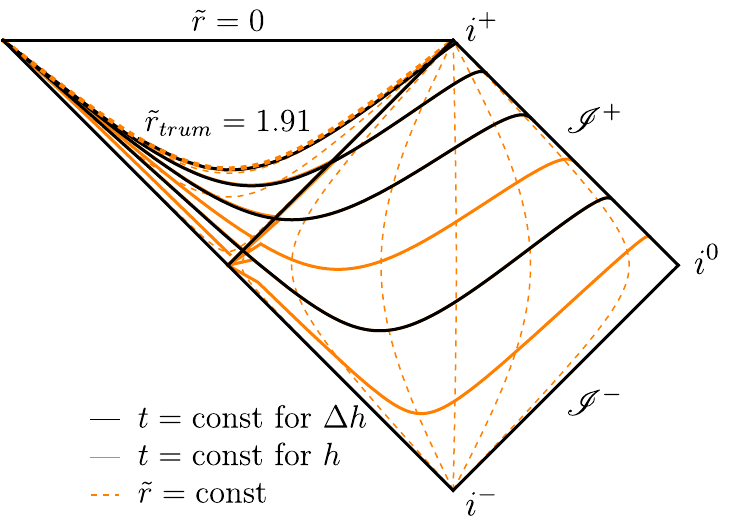}
\caption{Carter-Penrose diagrams of hyperboloidal CMC Schwarzschild trumpet slices. The black solid lines labelled as ``$t=$ const for [integration of] $\Delta h$'' are those constructed with the height function in \fref{heightpdh} integrating the smooth $\Delta h$, while the orange solid ones (``$t=$ const for [integration of] $h$'') correspond to using the height function in \fref{heightp}. The thin dashed orange lines denote hypersurfaces of constant radial coordinate, and the thick dashed orange line denotes the location of the trumpet, the innermost value of the Schwarzschild radial coordinate reached by these hyperboloidal CMC trumpet slices.}\label{penphysboth}
\end{figure}

For the construction of the Carter-Penrose diagram it is convenient to substitute the explicit expression of $f$ \eref{fksexpr} into \eref{TRSchw}, yielding the simpler expressions (valid inside and outside of the horizon)
\begin{subequations}
\begin{eqnarray}\label{RTdhp}
\tilde R&=&\frac{1}{2}\left[e^{\frac{t+\Delta h+\rtilde}{4M}}+\left({\frac{\rtilde}{2M}-1}\right)e^{-\frac{t+\Delta h-\rtilde}{4M}}\right] , \\  
\tilde T&=&\frac{1}{2}\left[e^{\frac{t+\Delta h+\rtilde}{4M}}-\left({\frac{\rtilde}{2M}-1}\right)e^{-\frac{t+\Delta h-\rtilde}{4M}}\right] . 
\end{eqnarray}
\end{subequations}
The result of plotting this as a conformal diagram is shown in the black lines of \fref{penphysboth}, which are compared to the orange ones depicting the result of integrating $h'(\rtilde)$ directly as done in the previous subsection. As the integration constants have been set in the same way, the spacetime slices coincide for the same chosen values of $t=$ const in both constructions. In the lower part of the diagram, some of the $\Delta h$ slices have been skipped to clearly illustrate the differences in behaviour around the horizon if the integration has been performed for the divergent $h$ or the smooth $\Delta h$. 

Given that time and radius are compactified in conformal diagrams, it makes no visual difference to plot the quantities up to $\rtilde\to\infty$ or up to a large value of $\rtilde$. So in this sense it is not a problem not to be able to integrate the height function up to infinite radius. However, the data output from the hyperboloidal code are expressed in terms of a compactified radial coordinate, so it is useful to devise a way to construct the Penrose diagrams for this setup. 

\subsection{Height function in compactified domain}\label{hcomp}

The use of a compactified radial coordinate $r$ allows the integration up to infinity of the height function that determines the hyperboloidal slice, and it is well adapted to the code's output data. 
The setup in the numerical evolutions of this work uses conformal compactification \cite{PhysRevLett.10.66} to include \scrip in the integration domain. That is achieved in two interrelated steps performed on a hyperboloidal slice, the compactification of the radial coordinate \eref{compac} and the conformal rescaling of the metric \eref{conf}: 
\begin{subequations}\label{confcompac}
\begin{eqnarray}\label{compac}
 &&\rtilde =\frac{r}{\aconf(r)}, \\
 && ds^2 = \Omega^2 d\tilde s^2 \quad \Leftrightarrow \quad g_{ab}=\Omega^2\tilde g_{ab}. \label{conf}
\end{eqnarray}
\end{subequations}
The quantity $\aconf$ is the compactification factor and $\Omega$ is the conformal rescaling, while $ds^2$ denotes the rescaled line element, and $g_{ab}$ and $\tilde g_{ab}$ are respectively the conformally rescaled and the physical metrics. 
The conformal factor is chosen to be the positive ($\Kc<0$), time-independent function
\begin{equation}\label{ein:omega}
\Omega(r) = \left(-\Kc\right)\frac{\rscri^2-r^2}{6\, \rscri},
\end{equation}
with $\rscri=1$ chosen as the coordinate location of null infinity in the compactified radial coordinate $r\in(0,\rscri)$. The compactification factor $\aconf$ is set so that the spatial metric on the CMC slice is conformally flat explicitly (imposing $\gamma_{rr}=\gamma_{\theta\theta}\equiv 1$ in \eref{e:linel}). For more details of how $\aconf$ is numerically determined see subsection IV E in \cite{\papbh}. 
The essence of conformal compactification is that $\aconf$ and $\Omega$ must behave the same, or at least be proportional to each other, as $\rtilde\to\infty$ (see e.g.~figure 5 in \cite{\papbh}): the blow-up of the radial coordinate is balanced by the rescaling of the metric. 
Applying the change in \eref{compac} to \eref{ein:lielphysh} yields
\begin{subequations}
\begin{eqnarray} \label{ein:lielphyscc1}
d\tilde s^2 &=& -A\,dt^2-2A\,\partial_{\rtilde}h\frac{\aconf-r\,\aconf'}{\aconf^2}dt\,dr \nonumber\\&&+\frac{\left[1-\left(A\,\partial_{\rtilde}h\right)^2\right]}{A}\frac{(\aconf-r\,\aconf')^2}{\aconf^4}d r^2 + \frac{r^2}{\aconf^2} d\sigma^2 \\
&=&-A\,dt^2 -2A\partial_{r}h\,dt\, dr\nonumber\\&&+\frac{\frac{(\aconf-r\,\aconf')^2}{\aconf^4}-\left(A\partial_{r}h\right)^2}{A}dr^2+\frac{r^2}{\aconf^2} d\sigma^2, \label{ein:lielphyscc}
\end{eqnarray}
\end{subequations}
where $A=A(\rtilde(r))=A\left(\frac{r}{\aconf}\right)$, $\aconf'=\partial_r\aconf(r)$, and $h$ depends on the radial coordinate specified by the derivative, that is, $h(\rtilde)$ in \eref{ein:lielphyscc1} and $h(r)$ in \eref{ein:lielphyscc}. Both expressions are equivalent; choosing one or the other will depend on the desired dependence for $h$.
Rescaling the line element by the conformal factor as in \eref{conf} provides 
\begin{eqnarray}\label{fsthyp}
ds^2&=& -A\,\Omega^2dt^2-2\,A\,\Omega^2\partial_{r}h\,dt\,dr\nonumber\\&+&\frac{\Omega^2}{\aconf^2}\left[\frac{\left[\frac{(\aconf-r\,\aconf')^2}{\aconf^2}-\left(\aconf A \partial_{r}h\right)^2\right]}{A}d r^2 + r^2 d\sigma^2\right] ,
\end{eqnarray}
which is used as initial data for the numerical simulations. 
The spherically symmetric line element in the conformally compactified domain is 
\begin{equation}\label{gconfcomp}
ds^2 = g_{tt} dt^2 + 2 g_{tr}\,dt\,dr +  g_{rr}\, dr^2 +  g_{\theta\theta}\, r^2\, d\sigma^2 .
\end{equation}
The relations between its metric components and those of \eref{gphys} in the physical domain are given in \aref{apix}.
The decomposition used in the code corresponds to the usual one in the 3+1 decomposition of spacetime (see for instance \cite{Alcubierre}) with a conformal rescaling of the spatial metric as in the BSSN-OK formalism \cite{NOK,PhysRevD.52.5428,Baumgarte:1998te,Brown:2007nt}, which in spherical symmetry takes the form
\begin{eqnarray}\label{e:linel}
ds^2 =&& - \left(\alpha^2-\frac{\gamma_{rr}}{\chi}{\beta^r}^2\right) dt^2 + 2\frac{\gamma_{rr}}{\chi}\beta^r dt\,dr \nonumber \\ &&+  \frac{\gamma_{rr}}{\chi}\, dr^2 +  \frac{\gamma_{\theta\theta}}{\chi}\, r^2\, d\sigma^2 .
\end{eqnarray}
The quantity $\alpha$ is called the lapse and determines how close the spacelike slices are to each other in evolution; $\beta^r$ is the radial component of the shift, which controls changes in spatial coordinates; $\chi$ is the spatial conformal factor, and $\gamma_{rr}$ and $\gamma_{\theta\theta}$ are components of the conformally-rescaled spatial metric. 
From now on, the relations between metric components and quantities of interest will be given for both metric decompositions, or only in terms of \eref{e:linel}.  
By comparing \eref{fsthyp} to \eref{gconfcomp} and \eref{e:linel}, the expression for the derivative of the height function in terms of the conformally compactified metric components is found
\begin{equation}\label{hcinteg}
h'(r) \equiv \partial_rh(r)=\frac{g_{tr}}{g_{tt}}=-\frac{\frac{\gamma_{rr}}{\chi}\beta^r}{\alpha^2-\frac{\gamma_{rr}}{\chi}{\beta^r}^2}
\end{equation}
For completeness, the closed-form expression for the height function for CMC slicing \eref{hfunc} in compactified coordinates is 
\begin{eqnarray}\label{hfunccompact}
h'(r) &=& \partial_rh(r) = \frac{d\rtilde}{dr}\partial_{\rtilde} h(\rtilde) = \frac{\aconf-r\,\aconf'}{\aconf^2} \left[\partial_{\rtilde} h\right](r/\aconf)
\nonumber \\ &=& - \frac{\left(\frac{\aconf-r\,\aconf'}{\aconf^2}\right)\left(\frac{\Kc\,r}{3\aconf} + \frac{\Cc\,\aconf^2}{r^2} \right)}{A\sqrt{A+\left(\frac{\Kc\,r}{3\aconf} + \frac{\Cc\,\aconf^2}{r^2}\right) ^2}},
\end{eqnarray}
where again $A=A\left(\frac{r}{\aconf}\right)=1-\frac{2M\aconf}{r}$ and $\aconf'=\partial_r\aconf(r)$. 

\subsubsection{Integrating $h$ in compactified radial coordinate $r$}

\begin{figure}
\center
\includegraphics[width=0.48\linewidth]{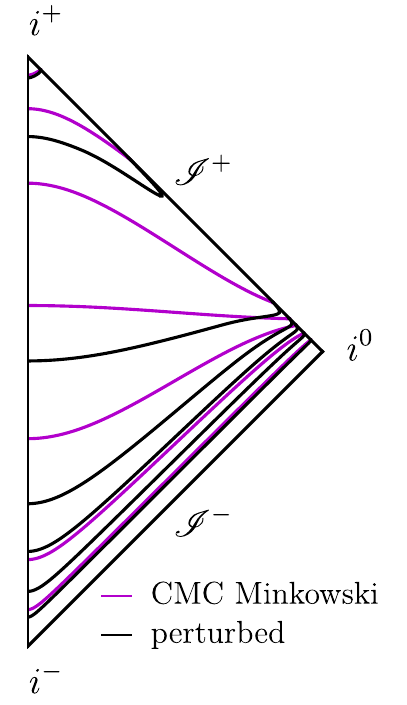}
\caption{Purple lines are CMC slices of Minkowski, while black ones correspond to perturbations due to a massless scalar field. The slices have been made to coincide at \scrip and thus the difference due to the presence of the scalar field is clearly seen in the interior of the domain. The deformation of the black slices by the scalar field's energy makes them timelike close to (but before reaching) \scrip, so that technically they no longer correspond to fully-spacelike (except at \scrip) hyperboloidal slices.}\label{minkcmcpert}
\end{figure}

Let us again consider the case of a regular spacetime to begin with. In terms of the compactified radial coordinate $r$, the CMC Minkowski height function \eref{hminkp} becomes
\begin{equation}\label{hminkc}
h(r) =\sqrt{\left(\frac{3}{\Kc}\right)^2+\left(\frac{r}{\Omega}\right)^2} + \frac{3}{\Kc}, 
\end{equation} 
where $\aconf$ has been substituted by $\Omega$, as is suitable in the absence of black holes. 
Performing the same radial transformation in \eref{basictrafo} together with the time change \eref{ttrafo} yields
{\small
\begin{subequations}\label{minkinidata}
\begin{eqnarray}
R_0=\frac{1}{2}\left[\arctan\left(t_0+h+\frac{r}{\Omega}\right)-\arctan\left(t_0+h-\frac{r}{\Omega}\right)\right],&& \\
T_0=\frac{1}{2}\left[\arctan\left(t_0+h+\frac{r}{\Omega}\right)+\arctan\left(t_0+h-\frac{r}{\Omega}\right)\right],&&
\end{eqnarray}
\end{subequations}
}
where the subscript ${}_0$ denotes the slice corresponding to a specific hyperboloidal time $t_0$. Substitution of \eref{hminkc} above is all there is left to create the Penrose diagram of Minkowski spacetime sliced along CMC \scrip-intersecting hyperboloidal slices. The result is given by the purple lines in \fref{minkcmcpert}. They all extend between $\rtilde=0$ ($r=0$) on the left vertical line to \scrip ($r=1$). 

The height function given by \eref{hcinteg} can also be integrated from numerical data, by providing the profiles for $\alpha$, $\beta^r$, $\gamma_{rr}$ and $\chi$. As an example, initial data of Minkowski plus a Gaussian-like perturbation of a massless scalar field with amplitude $\sim 0.07$ (the same used in the collapse simulations in \ssref{colevol}) is shown in the black lines in \fref{minkcmcpert}. The integration constant for the height function has been chosen so that the flat and perturbed slices intersect \scrip at the same locations. 

\begin{figure}
\center
\includegraphics[width=0.96\linewidth]{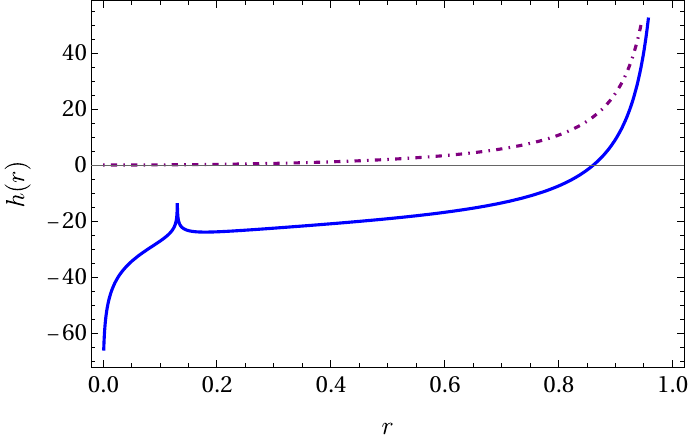}
\includegraphics[width=0.96\linewidth]{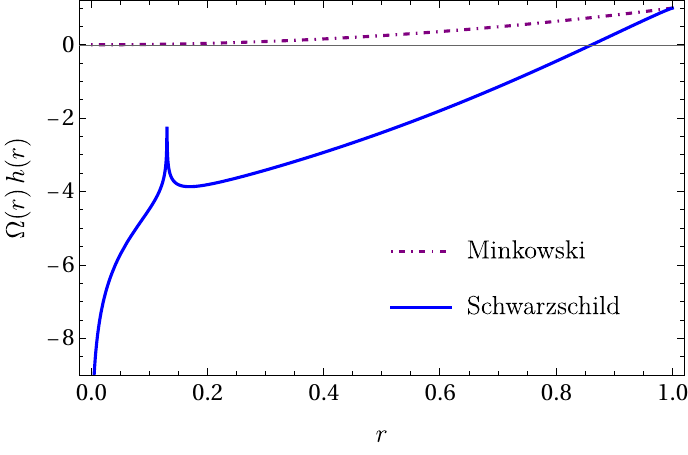}
\caption{The hyperboloidal height function expressed in terms of the compactified radial coordinate $r$ diverges as $r\to\rscri=1$ -- see top plot, where the Minkowski (purple dot-dashed line) and Schwarzschild (blue line) CMC height functions are depicted. In the bottom plot, $h$ is displayed multiplied by the conformal factor $\Omega$ \eref{ein:omega}, so that a finite value is attained at \scrip. The Schwarzschild height function diverges at the horizon in an equivalent way to the height function in \fref{heightp}.}\label{heightc}
\end{figure}

The result of integrating \eref{hfunccompact} for the height function in Schwarzschild CMC slices is shown in the top plot in \fref{heightc}, where \eref{hminkc} is also included for comparison. The integration takes place in two different domains of the compactified radial coordinate, separated by the locations where the height function diverges: between the trumpet ($\rtilde=\pr\trum$, $r=0$) and the horizon ($\rtilde=2M$, $r=r_{hor}\approx 0.1305$ for the chosen parameters), and between the horizon and \scrip ($\rtilde=\infty$, $r=\rscri=1$). Given that the expressions diverge at all those three points, at the numerical level the integration takes place up to almost the exact values.
The integration constant can be set to any convenient value, as that just adds or subtracts from the hyperboloidal time $t$. As done before, the integration constant is set so that the Schwarzschild height function coincides with the Minkowski one \eref{hminkc} for $r\to\rscri$. 
Both diverge at \scrip, as asymptotically the hyperboloidal slices become tangent to the null directions. However, $\Omega\,h$ is finite there, as can be seen in the bottom plot in \fref{heightc}. Namely, $\Omega\,h|_{\scri^+}=\rscri$,\footnote{This can be deduced from \eref{basic1} for Minkowski (it is the same in presence of a black hole): $\tilde U \equiv \tilde u = \tilde t-\rtilde = t + h - r/\Omega$. Asymptotically at \scrip, the hyperboloidal time $t\equiv\tilde u$, so $h|_{\scri^+} = r/\Omega|_{\scri^+}$, $\Omega\,h|_{\scri^+} = \rscri$.} which is set to unity here.

The behaviour of the height function at null infinity suggests integrating for the rescaled height function $H=\Omega\,h$ instead of the raw $h$. Note that while $h'(r)$ also diverges at \scrip, $\Omega^2h'$ is finite there. This procedure would involve solving for $H(r)$ the ordinary differential equation 
\begin{equation}
H'(r) = H(r)\frac{\Omega'}{\Omega}-\Omega\,h'(r), 
\end{equation}
where $h'(r)$ is known from \eref{hcinteg}. This differential equation is formally singular at \scrip, so that even if its solution $H(r)$ is finite there, it can pose trouble when solving numerically. No clear advantage was found when solving for $H$ instead of $h$ in the basic numerical tests performed for this work, so for simplicity the calculations are performed in terms of the original $h$.

The Schwarzschild equivalent to \eref{minkinidata} is the following

\begin{subequations}\label{schwinidata}
$\rtilde>2M$:
\begin{eqnarray}
R_0&=&\frac{1}{2}\left[\arctan\left(\sqrt{1-\frac{r}{2M\aconf}}\,e^{\frac{t_0+h(r)+\frac{r}{\aconf}}{4M}}\right)\right. \nonumber \\ && \left.-\arctan\left(\sqrt{1-\frac{r}{2M\aconf}}\,e^{-\frac{t_0+h(\tilde r)-\frac{r}{\aconf}}{4M}}\right)\right] ,\\
T_0&=&\frac{1}{2}\left[\arctan\left(\sqrt{1-\frac{r}{2M\aconf}}\,e^{\frac{t_0+h(r)+\frac{r}{\aconf}}{4M}}\right)\right. \nonumber \\ && \left.+\arctan\left(\sqrt{1-\frac{r}{2M\aconf}}\,e^{-\frac{t_0+h(\tilde r)-\frac{r}{\aconf}}{4M}}\right)\right] ,
\end{eqnarray}
$\rtilde<2M$: 
\begin{eqnarray}
R_0&=&\frac{1}{2}\left[\arctan\left(\sqrt{\frac{r}{2M\aconf}-1}\,e^{\frac{t_0+h(r)+\frac{r}{\aconf}}{4M}}\right)\right. \nonumber \\ && \left.+\arctan\left(\sqrt{\frac{r}{2M\aconf}-1}\,e^{-\frac{t_0+h(\tilde r)-\frac{r}{\aconf}}{4M}}\right)\right] ,\\
T_0&=&\frac{1}{2}\left[\arctan\left(\sqrt{\frac{r}{2M\aconf}-1}\,e^{\frac{t_0+h(r)+\frac{r}{\aconf}}{4M}}\right)\right. \nonumber \\ && \left.-\arctan\left(\sqrt{\frac{r}{2M\aconf}-1}\,e^{-\frac{t_0+h(\tilde r)-\frac{r}{\aconf}}{4M}}\right)\right] .
\end{eqnarray}
\end{subequations}
It is used to construct the conformal diagram after inserting the height function described above: the result looks basically like the black solid lines in \fref{penphysboth}, except for some minor imperfections around the horizon (looking like a much improved version of the orange line). The expressions above are also used as initial data for $R$ and $T$ in the trumpet evolutions in \ssref{trumpevol}. 

One way to improve the quality of the data around the horizon is integrating the height function using the compactified tortoise radius as done in  \aref{compt}. Better even is to take away the singularity as done in the physical domain in \ssref{hcalcpdh}. 

\subsubsection{Integrating $\Delta h$ in compactified radial coordinate $r$}

The expression for the derivative of the height function $f$ needs to account for the change to the compactified radial coordinate and the conformal rescaling. The conformally compactified equivalent to \eref{fprphysexpr} is 
\begin{eqnarray}\label{fprexpr}
f'(r) &=& -\frac{1-A}{A}\,\Omega^2\left(\frac{\aconf-r\aconf'}{\aconf^2}\right) = \frac{\Omega^2+g_{tt}}{g_{tt}}\left(\frac{\aconf-r\aconf'}{\aconf^2}\right) \nonumber \\
&=& \left(1-\frac{\Omega^2}{\alpha^2-\frac{\gamma_{rr}}{\chi}{\beta^r}^2}\right)\left(\frac{\aconf-r\aconf'}{\aconf^2}\right) , 
\end{eqnarray}
where $A=-g_{tt}/\Omega^2$ from comparing the first terms in \eref{fsthyp} and \eref{gconfcomp} was used. The compactification factor $\aconf$ is either determined numerically for a CMC slice following subsection IV E in \cite{\papbh} or is set from $\aconf=\Omega\sqrt{\frac{\chi}{\gamma_{\theta\theta}}}$, which follows from \eref{rel3} in the next subsection. 
The expression to integrate $\Delta h$ from is obtained by substituting \eref{hcinteg} and \eref{fprexpr} 
\begin{eqnarray}\label{dhcinteg}
\Delta h'(r) &\equiv& \partial_r\Delta h(r) = h'(r) - f'(r) \nonumber \\
&=&\frac{g_{tr}}{g_{tt}}-\frac{\Omega^2+g_{tt}}{g_{tt}}\left(\frac{\aconf-r\aconf'}{\aconf^2}\right) \\
&=&\frac{\left[\Omega^2 -\left(\alpha^2-\frac{\gamma_{rr}}{\chi}{\beta^r}^2\right)\right]\left(\frac{\aconf-r\aconf'}{\aconf^2}\right) -\frac{\gamma_{rr}}{\chi}\beta^r}{\alpha^2-\frac{\gamma_{rr}}{\chi}{\beta^r}^2}.   \nonumber
\end{eqnarray}
Even if this quantity is smooth and finite in the interior of the integration domain ($r\in(0,1)$), care may be required at the horizon when integrating numerically. 
\begin{figure}
\center
\includegraphics[width=0.96\linewidth]{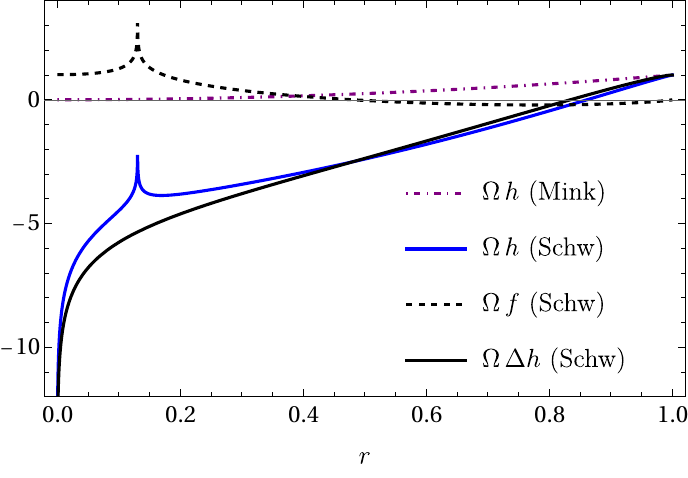}
\includegraphics[width=0.96\linewidth]{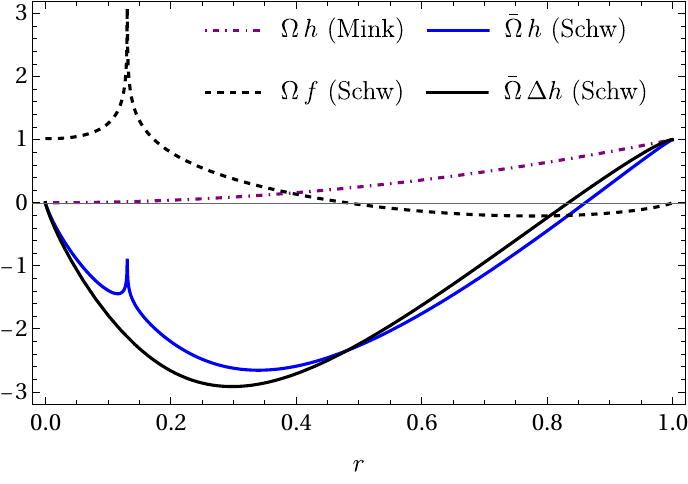}
\caption{The top plot displays $h(r)$ in blue, $\Delta h(r)$ in black, $f(r)$ in black dashed and, for comparison, the Minkowski height function, all of them rescaled by the conformal factor $\Omega$. The bottom plot displays $h$ and $\Delta h$ rescaled by the compactification factor $\aconf$ instead. It vanishes at $r=0$ fast enough to overpower the divergence of the height function at the trumpet.}\label{heightcdh}
\end{figure}
The results are shown in \fref{heightcdh}. This is the conformally compactified version of \fref{heightpdh}. The integrated $\Delta h(r)$ is displayed in black and the complete hyperboloidal height function $h$ in blue, coinciding with that in \fref{heightc}'s bottom plot. Two different rescalings are used for these quantities: in the top plot they are rescaled by $\Omega$, which counteracts the divergence of the height function at \scrip, while in the bottom plot they are rescaled by $\aconf$, whose fall-off at the origin is stronger than the height function's divergence. 

To construct the Penrose diagram, substitute $\Delta h(r)$ integrated from \eref{dhcinteg} into \eref{RTdhp} and substitute the physical radial coordinate there using \eref{compac}. The result will look the same as the black lines in \fref{penphysboth}. 

\subsection{Stationary Schwarzschild trumpet}\label{statio}

Spherically symmetric hyperboloidal CMC Schwarzschild trumpet initial data evolved with gauge conditions such as in \cite{\papgauge} attain a stationary state that corresponds to a trumpet geometry that is no longer CMC. The hyperboloidal time changes by a constant rescaling, similarly as in the transformation to hyperboloidal time (55) in \cite{PanossoMacedo:2019npm}. This rescaling is denoted here by $\cp$ (completely unrelated to the speed of light), as $t\to\cp\,t$. Its effect on the metric is that $g_{tt}$ gets rescaled by $\const$ and $g_{tr}$ by $\cp$. In the stationary regime achieved by the evolutions, the constant $\cp$ will take a different value depending on the gauge conditions chosen and the values of their parameters. See section V in \cite{\papbh} for more information. 

The quantity $\cp$ appears in the relations between metric components in the stationary regime. Whereas for CMC $\cp=1$, for the evolutions considered here the following expressions hold at stationarity (taken from (35) in \cite{\papbh}): 
\begin{subequations}\label{relsconf}
\begin{eqnarray}
-g_{tt}&=&\alpha^2-\frac{\gamma_{rr}{\beta^{r}}^2}{\chi}= 
 \const\,\Omega^2\left(1-\frac{2M\bar\Omega}{r}\right) \nonumber \\&=& \const\,\Omega^2\,A(r), \label{rel1}\\
 g_{rr}&=&\frac{\gamma_{rr}}{\chi}=\frac{\const}{\alpha^2}\left(\frac{\Omega^2}{\bar\Omega^2}(\bar\Omega-r\bar\Omega')\right)^2, \label{rel2}\\
 g_{\theta\theta}&=&\frac{\gamma_{\theta\theta}}{\chi}=\frac{\Omega^2}{\aconf^2}. \label{rel3}
\end{eqnarray}
\end{subequations}
with $\gamma_{\theta\theta}=\gamma_{rr}^{-1/2}$. This allows to find an expression for the compactification factor from \eref{rel3}, namely $\aconf=\Omega\sqrt{\chi\sqrt{\gamma_{rr}}}$. 
Knowing the value of $\cp$ is needed to build the Carter-Penrose diagram. It can be obtained from numerical metric data by isolating $\const$ from \eref{rel1} or \eref{rel2}, after substituting the expressions for the compactification factor, just described, and the conformal factor \eref{ein:omega}. 

Instead of using \eref{fprexpr} and \eref{dhcinteg}, now the expressions to calculate $f'(r)$ and $\Delta h'(r)$ include factors with the $\cp$ constant: 
\begin{eqnarray}\label{fpresc}
f'(r) &=& -\frac{1-A}{A}\,\Omega^2\left(\frac{\aconf-r\aconf'}{\aconf^2}\right) = \frac{\Omega^2+g_{tt}/\const}{g_{tt}/\const}\left(\frac{\aconf-r\aconf'}{\aconf^2}\right) \nonumber \\
&=& \left(1-\frac{\Omega^2\const}{\alpha^2-\frac{\gamma_{rr}}{\chi}{\beta^r}^2}\right)\left(\frac{\aconf-r\aconf'}{\aconf^2}\right) , 
\end{eqnarray}
\begin{eqnarray}\label{hpresc}
\Delta h'(r) &\equiv& \partial_r\Delta h(r) = h'(r) - f'(r) \nonumber \\
&=&\frac{g_{tr}/\cp}{g_{tt}/\const}-\frac{\Omega^2+g_{tt}/\const}{g_{tt}/\const}\left(\frac{\aconf-r\aconf'}{\aconf^2}\right) \\
&=&\frac{\left[\Omega^2\const -\left(\alpha^2-\frac{\gamma_{rr}}{\chi}{\beta^r}^2\right)\right]\left(\frac{\aconf-r\aconf'}{\aconf^2}\right) -\frac{\gamma_{rr}}{\chi}\beta^r\cp}{\alpha^2-\frac{\gamma_{rr}}{\chi}{\beta^r}^2}.   \nonumber
\end{eqnarray}

The result of constructing the height function for metric numerical data whose evolution has reached stationarity has not been included here, because it looks the same as the black solid lines in \fref{dyncomparpen}. The blue lines there correspond to solving for numerical data following the procedure in \ssref{compt} for CMC slices with $c=1$. 

\section{Dynamical spacetimes}\label{dyn}

In the case where the spacetime is not stationary or static, the construction with the height function from \sref{sta} is no longer valid. A method that does not rely on stationarity conditions is required. 

\subsection{Construction via eikonal equation}\label{eikonalbasics}

Here we build on the construction of causal diagrams described in the appendix of \cite{Thierfelder:2010dv} and in section III D in \cite{Hilditch:2016xzh}, based on the evolution of the eikonal equation. Originated in geometrical optics, here the eikonal equation will be used to demand that the initially null coordinates remain so during evolution. 
The starting point are the ingoing and outgoing null coordinates ($\tilde u,\tilde v$) as defined in \eref{ksl1} for the black hole spacetime, or as in \eref{basic1} ($\tilde U\equiv \tilde u = \tilde t-\rtilde$ and $\tilde V\equiv \tilde v = \tilde t+\rtilde$) for Minkowski. 
The eikonal equations are, 
\begin{subequations}
\begin{eqnarray}
\tilde g^{\tilde u\tilde u}&=&\tilde g^{ab}\tilde \nabla_a\tilde u\tilde \nabla_b\tilde u=0, \\
\tilde g^{\tilde v\tilde v}&=&\tilde g^{ab}\tilde \nabla_a\tilde v\tilde \nabla_b\tilde v=0 . 
\end{eqnarray}
\end{subequations}
Setting that $\tilde u$ and $\tilde v$ are functions of $t$ and $\rtilde$, and upon ansatz of the metric \eref{gphys} with $\tilde g_{tt}=-(\tilde\alpha^2-\tilde g_{\rtilde\rtilde}{\beta^{\rtilde}}^2)$ and $\tilde g_{t\rtilde}=\tilde g_{\rtilde\rtilde}{\beta^{\rtilde}}$, they decompose into the advection equations
\begin{eqnarray}
\begin{array}{l}\partial_t\tilde u=-\tilde c_+\partial_{\tilde r}\tilde u , \\  \partial_t\tilde v=-\tilde c_-\partial_{\tilde r}\tilde v ,\end{array} \quad\textrm{with}\quad \tilde c_\pm=\pm \frac{\tilde\alpha}{\sqrt{\tilde g_{\rtilde \rtilde}}}-\beta^{\tilde r} ,
\end{eqnarray}
where $\tilde c_\pm$ are the incoming and outgoing radial physical light speeds. 
The idea is to use these equations to evolve the slices in the conformal diagrams, which are given by $T$ and $R$ as in \eref{basic3}. Given that the code uses the compactified radial coordinate, the desired setup is to evolve $T(t,r)$ and $R(t,r)$. 
It is important to note that all coordinate transformations in \eref{basictrafo} and \eref{ksl} are either linear combinations of variables (\eref{basic1}, \eref{basic3}, \eref{ksl1} and \eref{ksl3}) or they involve a rescaling of separate quantities (\eref{basic2} and \eref{ksl2}). They preserve the form of the advection equation and allow to obtain for the compactified $U$ and $V$ 
\begin{equation}
\partial_tU=-\tilde c_+ \partial_{\rtilde} U \quad \textrm{and} \quad \partial_tV=-\tilde c_-\partial_{\tilde r}V . 
\end{equation}
Consider now the operator $\tilde c_\pm \partial_{\rtilde}$ acting on some field $X$, and use the transformation rule of the propagation speeds as given by \eref{speedtrafo} and the relation between $\rtilde$ and $r$ in \eref{compac}. The result yields
\begin{equation}
\tilde c_\pm \partial_{\rtilde} X = c_\pm \left(\frac{\aconf-r\aconf'}{\aconf^2}\right) \frac{dr}{d\rtilde} \partial_rX = c_\pm \partial_rX ,
\end{equation}
which lets us write
\begin{equation}
\partial_tU= - c_+ \partial_rU \quad \textrm{and} \quad \partial_tV=-c_-\partial_{r}V. 
\end{equation}
The actual evolution equations implemented are the following, obtained from those above by using \eref{basic3} and setting $c_\pm=\pm\alpha\sqrt{\frac{\chi}{\gamma_{rr}}}-\beta^r$: 
\begin{subequations}\label{eikonalrt}
\begin{eqnarray}
\partial_tR &=&\beta^r\partial_rR+\alpha\sqrt{\frac{\chi}{\gamma_{rr}}}\partial_rT, \\
\partial_tT &=&\beta^r\partial_rT+\alpha\sqrt{\frac{\chi}{\gamma_{rr}}}\partial_rR.  
\end{eqnarray}
\end{subequations}
This corresponds to (A6) in \cite{Thierfelder:2010dv} with the change of notation $X\to R$ and expressed in terms of the compactified and conformally rescaled metric components used in the present work. Also, in the equations above $t$ is the hyperboloidal time coordinate and $r$ is the compactified radial coordinate on the hyperboloidal slice. 

Minkowski initial data for $R$ and $T$ are given by \eref{minkinidata}, while \eref{schwinidata} provides hyperboloidal CMC trumpet initial slices. 
Pairs of $(R(t_0,r),T(t_0,r))$ are set for several values of $t_0$ and evolved in the code using \eref{eikonalrt}. To create the Penrose diagrams from the evolution data, $T(t,r)$ is plotted versus $R(t,r)$, which in practice translates to plotting the points given by the pairs $(R(t_n,r_i),T(t_n,r_i))$ (for all $n$ steps in the time evolution and all $i$ in the spatial grid) of the gridfunctions obtained as output from the code. 
This procedure is now applied to visualize hyperboloidal trumpet slices relaxing and the collapse into a black hole. 

\subsection{Evolution of trumpet slices}\label{trumpevol}

\begin{figure}
\center
\includegraphics[width=0.96\linewidth]{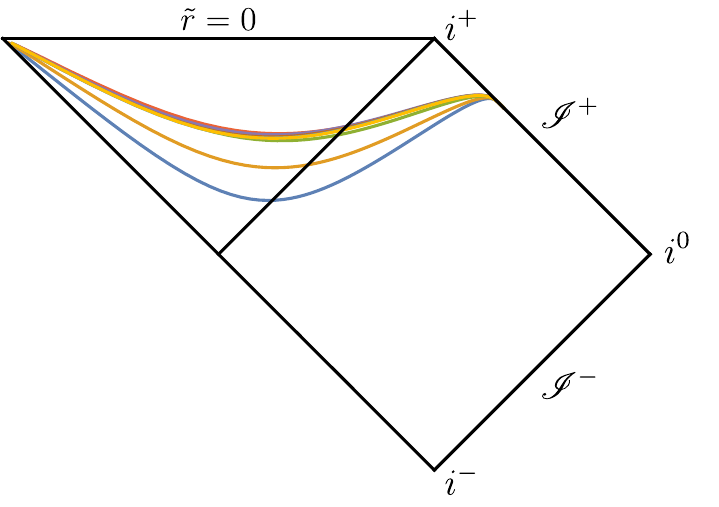}
\includegraphics[width=0.96\linewidth]{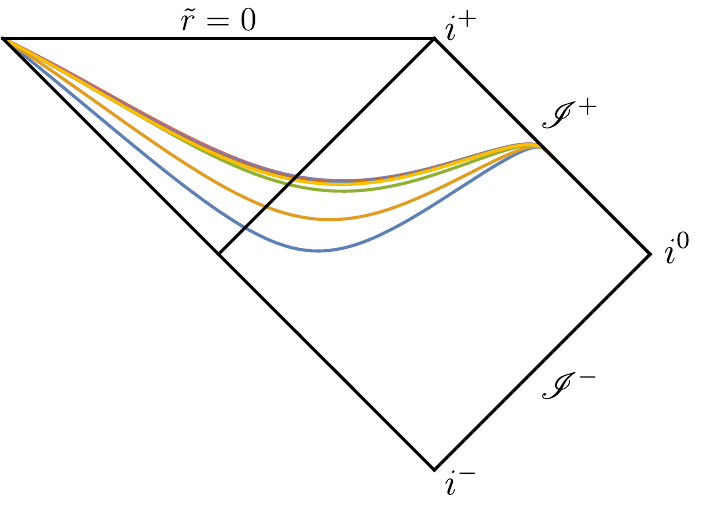}
\includegraphics[width=0.96\linewidth]{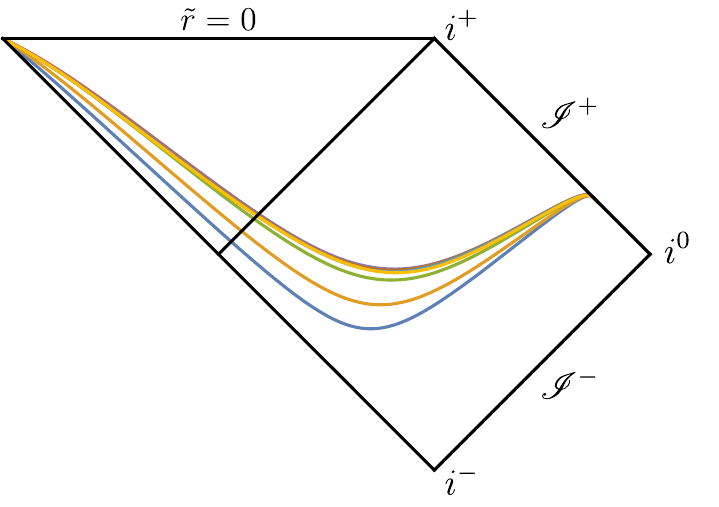}
\includegraphics[width=0.96\linewidth]{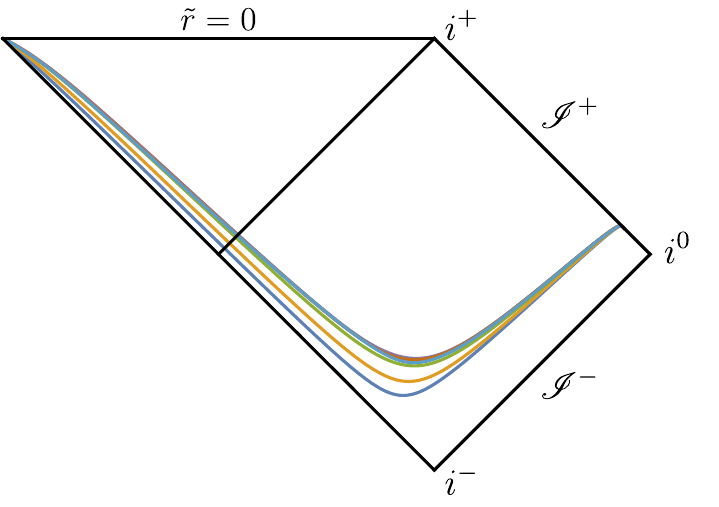}
\caption{Superposition of $(R,T)$ data corresponding to initial data for different initial $t_0$ at the times when they overlap at \scrip. This shows the gradual change in the trumpet slices.}\label{dyntrump}
\end{figure}

If CMC trumpet initial data is evolved with gauge conditions adapted to the hyperboloidal setup as described in \cite{\papgauge}, then there is some initial dynamics in the simulation where the CMC trumpet relaxes into a stationary solution of the gauge conditions. 
There is no change in the geometry of the spacetime, only gauge dynamics that induce a change in the profiles of the slices. 
This effect is shown on the diagrams in \fref{dyntrump}. 
CMC data given by \eref{schwinidata} set for different initial $t_0$ are evolved using \eref{eikonalrt} together with the Einstein equations with CMC trumpet initial data (see \cite{\papbh} for more details). Extrapolation boundary conditions are imposed for $R$ and $T$ at the trumpet ($r=0$) and at \scrip ($r=1$). This allows to obtain the profiles of the slices on the fly in the dynamical regime, and conformally compactified hyperboloidal data is well suited to cover the whole range of the slices (up to future null infinity). Visualization is completely straightforward, as simply the values of $R$ are to be plotted on the horizontal axis and those of $T$ on the vertical one, providing output diagrams as those in \fref{dyntrump}. 
The slices shown in each of the diagrams have been chosen in the following way: a slice built at $t_0$ is shown at its initial state ($t=0$, where t is the hyperboloidal time used by the code), then one built at $t_0-1$ is plotted at its $t=1$, one built at $t_0-2$ is depicted at its $t=2$, etc. The pattern is that $N$ slices built at $t_0-t_{N}$ are evolved for $t_{N}$ before being plotted. That ensures that the slices coincide at \scrip, while at the same time showing the change in a trumpet slice as time goes by, from its initial CMC profile to its final stationary state. Figure \ref{dyntrump} shows 4 sets of about $N=8$ each to illustrate the effect in different parts of the Penrose diagram. The time elapsed between slices is of $\Delta t=1$, so that after $t\sim 4$ of evolution, the stationary regime is already achieved on the conformal diagrams from a visual point of view. 

\begin{figure}
\center
\includegraphics[width=0.96\linewidth]{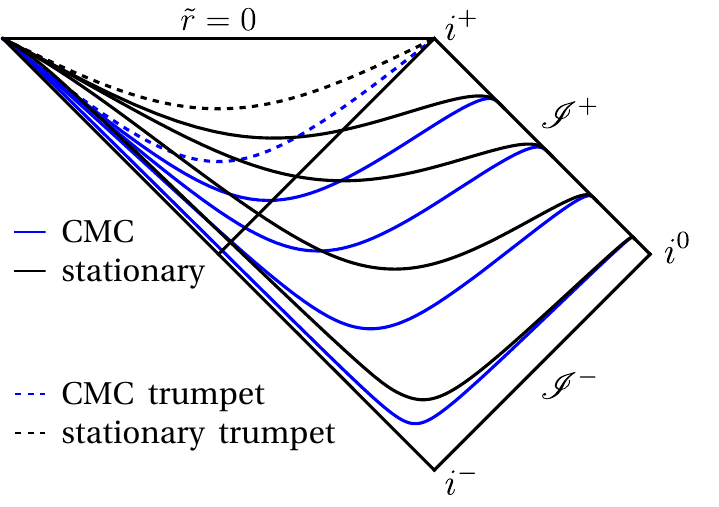}
\caption{Conformal diagram showing initial CMC slices and final stationary non-CMC ones from evolved $(R,T)$ data. The dashed lines denote the location of the trumpet in terms of the Schwarzschild radial coordinate. Note that it is closer to the singularity in the stationary case.}\label{dyncomparpen}
\end{figure}

Figure \eref{dyncomparpen} compares the initial CMC slices to the stationary ones found in the evolution, using data from the same simulations as in \fref{dyntrump} and employing the same method to make CMC slices at $t=0$ coincide at \scrip with (initially CMC) evolved slices evolved until the stationary regime has been attained. This plot is to be compared to figure 11b in \cite{\papbh}. The CMC slices (blue) in both Penrose diagrams have been constructed in the same way. However, the stationary non-CMC ones (in black) have been built differently. In \cite{\papbh}'s figure 11b, the height function approach with integration in terms of the compactified tortoise coordinate, as described in \ssref{compt} in this work, has been applied for non-CMC initial data determined in that work. Here the black lines correspond to the numerical evolution of the blue ones using \eref{eikonalrt}.

\begin{figure}
\center
\includegraphics[width=0.96\linewidth]{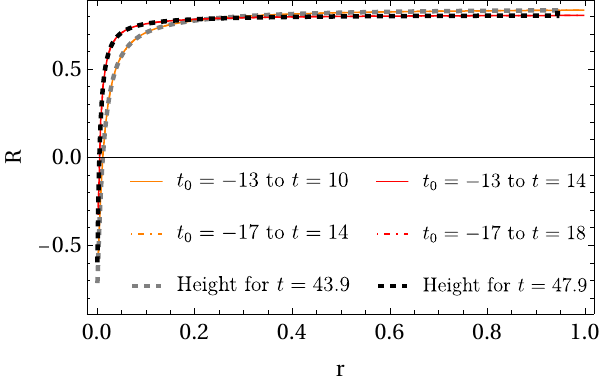}
\includegraphics[width=0.96\linewidth]{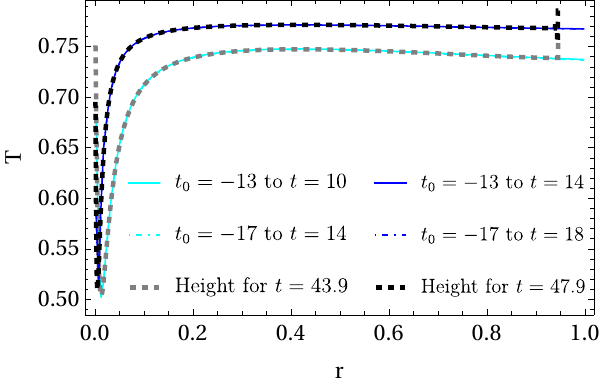}
\includegraphics[width=0.843\linewidth]{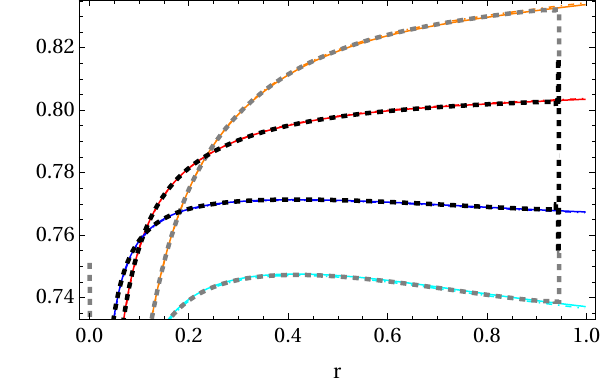}
\includegraphics[width=0.117\linewidth]{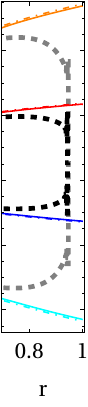}
\caption{Quantities $R$ (top plot) and $T$ (middle one) for two times separated by $\Delta t=4$, to show that the dynamical evolution (thin lines) coincides with the height function construction of the stationary data (thick dashed lines). The red (for $R$), blue (for $T$) and black (for both) profiles correspond to a later time than the orange, cyan and gray lines. The thin coloured dashed lines have been set at an earlier $t_0$ and thus evolved for longer than their solid-line counterparts. Both sets are included to show that the evolved data agree well in the stationary profiles. The bottom left plot shows $R$ and $T$ in more detail to better appreciate the good coincidence of the curves. The small bottom-right plot shows the equivalent to part of the left plot if the value $c=1$ is used in \eref{fpresc} and \eref{hpresc} when building the slices from the same numerical metric data: the coincidence is clearly worse. The thick dashed lines do not extend all the way to \scrip because of numerical errors in the construction.}\label{RTcompar}
\end{figure}

The final state of the evolved slices is compared to the height function construction described in \ssref{statio} in \fref{RTcompar}.
The comparison is showed for the quantities $R$ and $T$ as a function of $r$ instead as in a Penrose diagram, because the former allows to better appreciate any potential discrepancies. 
The profiles corresponding to height-function-built slices are denoted by thick dashed lines, gray for an earlier time in the foliation ($t=43.9$) and black for a later time ($t=47.9$). The time has been chosen so that the slice data coincides well with the numerically-evolved counterparts. The absolute time cannot be used for comparison, as a different integration constant for the height function alters it. Thus, the comparison is included for two different times (separated by $\Delta t=4$), showing that coincidence between the curves happens in both cases. To ensure reliability of the evolved data, two sets of data have been used for each time: one built at an earlier time and evolved for longer and one more recent (with corresponding initial times given in the legend). in the same way as done for \fref{dyntrump}, but here at a time late enough such that the stationary regime has been achieved. 
The factors of $\cp$ in \eref{fpresc} and \eref{hpresc}, as well as \eref{dtortresc} and \eref{hptortresc}, were checked making this comparisons: only the correct rescalings provide a good fit between height function and eikonal-evolved curves. The bottom plots illustrate this: the height-function-constructed thick dashed lines on the left one were calculated using the $\cp$ factors as \eref{fpresc} and \eref{hpresc}, whereas those on the narrow right plot used the same initial data but were built with $\cp=1$. Even if only the range $r\in[0.8,1]$ is shown for both configurations, it is clear that the factors of $\cp$ are required for coincidence. 
Note also that in the bottom plot in \fref{RTcompar} the coincidence between the height-function-determined curves (thick dashed lines) and the evolved $R$ and $T$ (thinner lines) is slightly better for the data at later time (black dashed as well as red and blue lines). This is because the system has relaxed more into the stationary state, and the assumption of stationarity that the construction in \ssref{statio} is based on is better respected\footnote{It is not possible to compare the profiles of $R$ and $T$ at very late times in a useful way. This is because of the effect of the shift $\beta^i$: it pushes the data outside of the horizon and towards future null infinity, in a similar way as shown in figure 15 in \cite{Hannam:2008sg}. This effect can also be seen in the upper slices of \fref{collapse}.}. 

\subsection{Evolution of collapse slices}\label{colevol}

In the just treated scenario of trumpet relaxation, the dynamics that takes place is not of physical nature, but due to a gauge effect: the geometry is Schwarzschild at all times, but the actual trumpet slice changes its profile. Trumpet relaxation reflects a gauge effect on a stationary spacetime. In the case of a collapse, there is a fundamental change in the physics of the system, as an initially regular setup transforms into a spacetime including a singularity. Here the simplest case of the collapse due to a massless scalar field in spherical symmetry is considered. The initial perturbation in the scalar field, with a Gaussian-like profile along the radial coordinate, is a shell that deforms Minkowski as in the black slices depicted in \fref{minkcmcpert}. 
The amplitude of the perturbation ($0.07$) is large enough so that the energy that gathers at $r=0$ when the scalar field reaches there creates a singularity. 

\begin{figure}
\center
\includegraphics[width=0.49\linewidth]{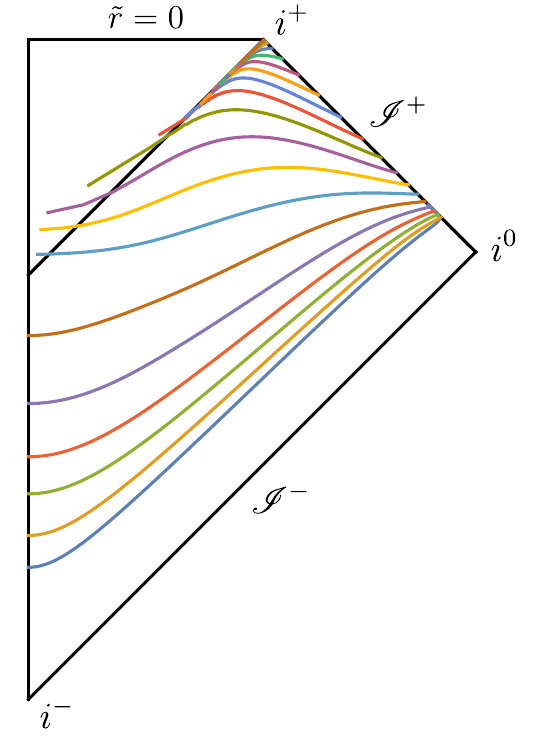}
\includegraphics[width=0.49\linewidth]{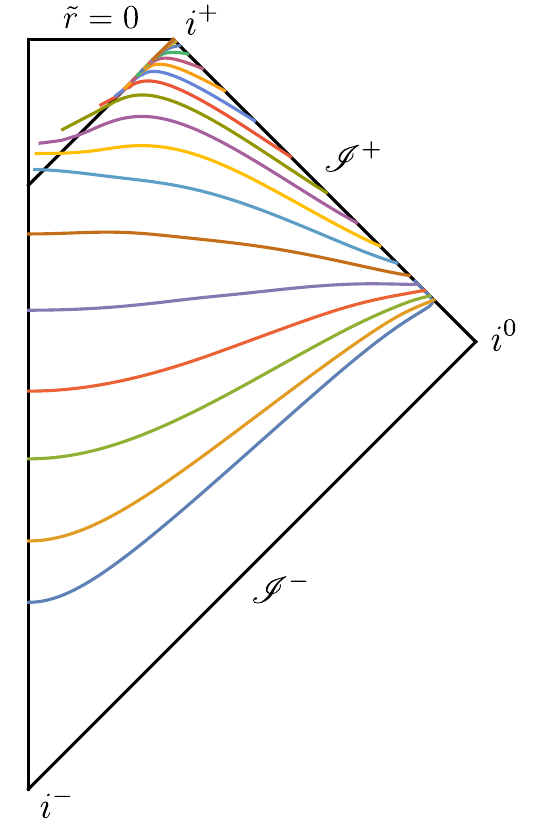}
\caption{Hyperboloidal slices corresponding to the collapse scenario. The initial data correspond to the foliation depicted in black in \fref{minkcmcpert}. On the left, the initial slice (lowest blue curve) was chosen $t_0=-0.7$ as in \eref{minkinidata} (with $h$ calculated from the metric perturbed by the scalar field), and on the right for $t_0=0$. The time interval between depicted slices is of $\Delta t=0.5$. The apparent horizon of the black hole (with final mass $\sim0.11$ in code units) forms at $t\approx 4.4$. Then, the value of $\tilde r\trum$, given by \eref{tildetrumdef} and zero for a regular spacetime, becomes $\sim0.15$, settling to $\sim0.16$ later in the simulation.}\label{collapse}
\end{figure}

The evolution of the hyperboloidal spacetime slices corresponding to this collapse scenario is shown in \fref{collapse}. The boundary conditions imposed at the origin are that $R$ is odd and $T$ is even there for as long as the spacetime is regular. Once collapse happens, extrapolation boundary conditions are used for those variables at $r=0$, in the same way as done in the trumpet relaxation evolution described in the previous subsection. Extrapolation boundary conditions are used at \scrip at all times. 

Both diagrams in \fref{collapse} represent the exact same physical scenario, but the time at which the initial slice for ($R,T$) was set is different. This is the reason why the $\rtilde=0$ black horizontal line representing the singularity, the black diagonal line representing the event horizon and that representing \scrip have different lengths in the diagrams. What is the same is the time elapsed since the initial slice and the collapse, even if it looks deformed due to the conformal compactification of the Carter-Penrose diagrams. The ``cover'' of the diagrams, meaning the black lines that delimit the spacetime and the horizon, has been adapted in such a way that the piling slices at late times coincide with the horizon joining future timelike infinity $i^+$. The evolution is carried out for a finite time, while $i^+$ corresponds to $t=\infty$. However, the coordinate transformations used in the construction of Penrose diagrams make it appear as if the later slices were reaching future timelike infinity. 

\section{Conclusions}\label{conclu}

Hyperboloidal evolutions of strong field data in spherical symmetry motivated the use of Carter-Penrose diagrams for visualization of the slices used in the simulations. Hyperboloidal slices reach future null infinity, an ingoing null surface for asymptotically flat spacetimes that does not allow any physical information to enter, and is thus very useful as outer boundary of the integration domain. 
The study of these slices has led to the work presented here. 
After illustrating the basic construction of conformal diagrams, the focus was put on the integration of the height function for hyperboloidal trumpet foliations of the Schwarzschild spacetime, for both closed-form CMC expressions and for numerical metric data. 
Also, a way to easily evolve ready-to-plot data during the dynamical regime was developed, based on the eikonal equation and previous work in the literature. The elegant Penrose diagrams produced are useful for understanding the properties and behaviour of the relaxing trumpet foliation and the evolution of the collapsing slices. 

Further work on this line will consider the Reissner-Nordstr\"om spacetime, describing a spherically symmetric black hole. Its stationary CMC trumpet height-function-constructed diagrams are provided in figures 3.11-14 in \cite{\alexthesis}. Once hyperboloidal simulations for this spacetime become available, its dynamical regime will be studied and illustrated as conformal diagrams following this approach.

\section*{Acknowledgements}

The author thanks Edgar Gasper\'in, An{\i}l Zengino\u{g}lu, David Hilditch and Christian Peterson-B\'orquez for valuable comments on the manuscript. 
The author is also indebted to David Hilditch for fruitful discussions on the eikonal construction and to Edgar Gasper\'in for useful references on conformal diagrams. 
The author thanks the Fundac\~ao para a  Ci\^encia e Tecnologia (FCT), Portugal, for the financial support to the Center for Astrophysics and Gravitation (CENTRA/IST/ULisboa) through the Grant Project~No.~UIDB/00099/2020. 
Acknowledged is also the financial support by the EU’s H2020 ERC Advanced Grant Gravitas–101052587, and that from the VILLUM Foundation (grant VIL37766) and the DNRF Chair program (grant DNRF162) by the Danish National Research Foundation. This project received funding from EU's Horizon 2020 MSCA grants 101007855 and 101131233.

\appendix
\section{Relation between physical and conformally compactified quantities} \label{apix}

The compactification of the radial coordinate is given by \eref{compac}, while the conformal rescaling of the metric is \eref{conf}. 
The values of the compactified radius used in the code are $r\in[0,\rscri]$ (or rather $r\in(0,\rscri)$ if a staggered grid is used), while the physical Schwarzschild radial coordinate takes values $\pr\in[\pr\trum,\infty)$. The value $\pr\trum$ (mapped to $r=0$ in the compactified radial coordinate, and first introduced in \ssref{hcalcp}) corresponds to the location of the trumpet throat and is calculated by 
\begin{equation}\label{tildetrumdef}
\pr\trum=\lim_{r\to 0}\frac{r}{\aconf(r)}. 
\end{equation}
This can be deduced from \eref{compac}. In the CMC case, $\pr\trum$ corresponds to the double root of the square root in eq. (7) in \cite{Baumgarte:2007ht}, or equivalently in eq. (3.40) in \cite{\alexthesis} and (22) in \cite{\papbh}, where $\pr\trum$ is called $R_0$.

The metric quantities in the uncompactified physical domain (on the left) are related to their conformally compactified equivalents as\footnote{These relations correspond to (28) and (29) in \cite{\papbh}, where $g$ is called $X$.}
\begin{subequations}\label{metphysconftrafos}
\begin{eqnarray}
\tilde\alpha(\pr) &=& \frac{\alpha}{\Omega} , \\
\tilde\chi(\pr) &=& \chi\,\frac{\Omega^2}{\aconf^2} , \\
\beta^{\pr}(\pr) &=& \beta^r\frac{\aconf-r\partial_r\aconf}{\aconf^2} , \\
\tilde \gamma_{\pr\pr}(\pr) &=& \gamma_{rr}\left(\frac{\aconf}{\aconf-r\partial_r\aconf}\right)^2 , \\
\tilde g_{\pr\pr}(\pr) &=& \frac{\tilde \gamma_{\pr\pr}(\pr)}{\tilde\chi(\pr)} = \frac{g_{rr}}{\Omega^2}\left(\frac{\aconf^2}{\aconf-r\partial_r\aconf}\right)^2 , \\
\tilde g_{\theta\theta}(\pr) &=& \frac{\tilde \gamma_{\theta\theta}(\pr)}{\tilde\chi(\pr)} =  g_{\theta\theta}\,\frac{\aconf^2}{\Omega^2} , \quad \tilde \gamma_{\theta\theta}(\pr) = \gamma_{\theta\theta}, 
\end{eqnarray}
\end{subequations}
where all quantities in the right-hand-sides are functions of the compactified radius $r$.
The compactification factor is calculated from $\aconf=\Omega\sqrt{\frac{\chi}{\gamma_{\theta\theta}}}$, with $\gamma_{\theta\theta} = \gamma_{rr}^{-1/2}$ and all quantities functions of $r$. Solving for $r$ in \eref{compac} after having set the just given expression for $\aconf$ provides the required relation $r(\pr)$ to substitute $r$ in \eref{metphysconftrafos}, so allowing to express all metric quantities in terms on the uncompactified radial coordinate. 

The relations holding in the stationary regime in the physical domain, equivalent to \eref{relsconf}, are
\begin{subequations}\label{relsphys}
\begin{eqnarray}
&&-\tilde g_{rr} = \tilde \alpha^2-{\beta^{\tilde r}}^2\frac{\tilde\gamma_{\tilde r \tilde r}}{\tilde\chi} = 
\const\left(1-\frac{2M}{\tilde r}\right), \\ &&
\tilde g_{\tilde r \tilde r}=\frac{\tilde \gamma_{\tilde r \tilde r}}{\tilde\chi}= \frac{\const}{\tilde\alpha^2}, \qquad 
\tilde g_{\theta\theta}=\frac{\tilde \gamma_{\theta\theta}}{\tilde \chi}= 1. \label{relsphysX}
\end{eqnarray}
\end{subequations}
They incorporate the $\cp$ factor that was not included in \cite{\papbh}'s equation (33) .

The relations between characteristic speeds in the physical and conformally compactified domains (used in \ssref{eikonalbasics}) are the following. 
From relations (28) and (29) from \cite{\papbh}, the propagation speeds associated to the physical and conformally compactified domains are respectively
\begin{equation}
\tilde{c}_{\pm} = \pm \frac{\tilde\alpha}{\sqrt{\tilde{g}_{\rtilde\rtilde}}}-\beta^{\rtilde} \quad \textrm{and} \quad
c_{\pm} = \pm \frac{\alpha}{\sqrt{g_{rr}}}-\beta^r . 
\end{equation}
The transformation rule between both is obtained by substituting the rules for the metric elements specified above and yields
\begin{equation}\label{speedtrafo}
\tilde{c}_{\pm} = \left(\frac{\aconf-r\aconf'}{\aconf^2}\right) c_{\pm}  . 
\end{equation}

\section{Calculation of height function via tortoise coordinate} \label{apextort}

The use of the tortoise coordinate as defined in \eref{dtortdr} allows to better deal with the coordinate singularity at the horizon: it does not take the singularity away, but brings it to infinity. Thanks to that, the height function can be integrated more accurately around the horizon. 

\subsection{Integrating $h$ in physical tortoise radial coordinate $\tort$}

The tortoise radial coordinate $\tort$ moves the coordinate singularity of the horizon to $-\infty$, at the price of having to deal with the interior and exterior of the black hole separately, see \eref{ptortexpr} and \eref{TRSchw}. This allows a more accurate numerical integration of the height function at the horizon. 

\begin{figure}[ht]
\center
\includegraphics[width=0.96\linewidth]{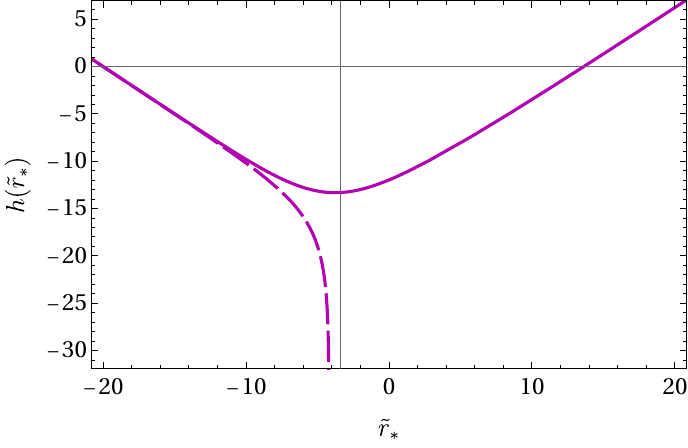}
\caption{Hyperboloidal height function expressed in terms of the physical tortoise radial coordinate. The solid line corresponds to the part of the height function outside of the horizon, while the dashed one is located inside. If expressed in terms of the Schwarzschild radial coordinate $\rtilde$ instead, the height function would look like \fref{heightp}, up to a different value of the integration constant. The use of the tortoise coordinate provides reliable data much closer to the horizon (on the left).}\label{heightpt}
\end{figure}

From a practical point of view, it may be useful to express the physical Schwarzschild radial coordinate in terms of the tortoise one. Obtaining $\rtilde(\tort)$ can be done numerically. 
The first derivative of height function in terms of $\tort$ is given by 
\begin{equation}\label{hptinteg}
h'(\tort)\equiv\partial_{\tort} h(\tort) = A \partial_{\pr} h(\pr) = -\tilde g_{t\rtilde}(\rtilde), 
\end{equation}
where in the first equality the chain rule with substitution of \eref{dtortdr} has been used. The right-hand-side is obtained by comparing the mixed terms of \eref{ein:lielphysh} and \eref{gphys}, or equivalently setting $\tilde g_{tt}=-A$ in \eref{hpinteg} and reorganising. 
The quantity $A\,h'(\pr)$ is a natural object in spherically symmetric spacetimes with good properties wherever $A$ vanishes, including cosmological horizons. 
Integrate 
\begin{equation}
\partial_{\tort} h(\tort) = -\tilde g_{t\rtilde}(\rtilde(\tort))
\end{equation}
separately for $\tort\in(-\infty,\infty)$ outside of the horizon, and for $\tort\in(-\infty,\tort{}\trum)$ inside of it. Here $\tort{}\trum$ denotes the value of the tortoise coordinate corresponding to the location of the trumpet, which in the Schwarzschild radial coordinate is denoted as $\pr\trum$ as was mentioned earlier. Thus $\tort(\pr\trum)=\tort{}\trum$. The integration is to be performed as
\begin{subequations}\label{integptort}
\begin{eqnarray}
&\pr>2M: \ h_{out}(\tort) = -\int_{-\infty}^{\infty}\tilde g_{t\rtilde}(\rtilde_{out}(\tort))d\tort, \label{integptortout}\\
&\pr<2M: \ h_{in}(\tort) = -\int_{-\infty}^{\tort{}\trum}\tilde g_{t\rtilde}(\rtilde_{in}(\tort))d\tort. \label{integptortin}
\end{eqnarray}
\end{subequations}
In practice it is not possible to integrate numerically to/from infinity. For the present visualization purposes it is enough to consider the approximate integration limits $\tort\in(-40,100)$ outside and $\tort\in(-40,\tort{}\trum)$ inside of the horizon. The result for the hyperboloidal CMC slice under consideration is shown in \fref{heightpt}. 
The final conformal diagram is created by substituting the results of \eref{integptortout} and \eref{integptortin} (via the change to hyperboloidal time \eref{ttrafo}) into \eref{ed:TRSchwout} and \eref{ed:TRSchwin} respectively. Then choose to express the quantities as functions of $\rtilde$ by using \eref{ptortexpr}, or as functions of $\tort$ by substituting all $\rtilde$ in terms of $\tort$ (inverting \eref{ptortexpr}). This is actually made more straightforward by starting from \eref{ed:TR} (already expressed in terms of the tortoise coordinate) and substituting \eref{integptort} there. 
Finally, $T(t,\tort)$ is plotted versus $R(t,\tort)$. The result appears like the black solid lines in \fref{penphysboth}, which look continuous and smooth around the horizon. Even if the coordinate singularity still exists at the horizon, the use of the tortoise coordinate is a clear improvement with respect to integrating the height function using the physical Schwarzschild radial coordinate (orange slices). 

\subsection{Integrating $h$ in compactified tortoise radial coordinate $\torc$} \label{compt}

A combination of the compactified radius together with the effect at the horizon of the tortoise radial coordinate motivates the introduction of a compactified tortoise coordinate $\torc$, as it provides a better treatment both at the horizon and at \scrip. 

\begin{figure}[ht]
\center
\includegraphics[width=0.96\linewidth]{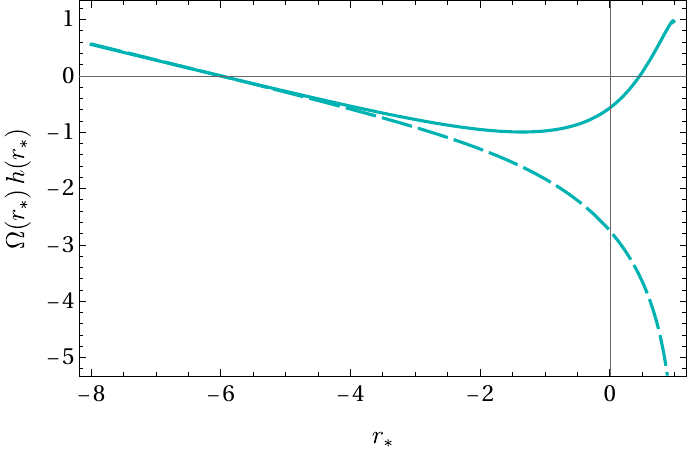}
\caption{The conformal factor $\Omega$ times the hyperboloidal height function expressed in terms of the tortoise compactified radial coordinate. It is equivalent to \fref{heightpt}, with the exception that the values of the compactified tortoise coordinate are bounded at the trumpet and at \scrip.}\label{heightct}
\end{figure}

The relation to determine the compactified tortoise coordinate, based on \eref{dtortdr} and including the value of the parameter $\cp$, is now written in terms of the conformally rescaled metric components 
\begin{equation}\label{dtortresc}
\frac{d\torc}{dr} = \frac{1}{A} = -\frac{\Omega^2\cp^2}{g_{tt}} =  \frac{\Omega^2\cp^2}{\alpha^2-\frac{\gamma_{rr}{\beta^{r}}^2}{\chi}}, 
\end{equation}
where the first equality used the equivalence $A=-g_{tt}/(\Omega^2\cp^2)$ from comparing the first terms in \eref{fsthyp} and \eref{gconfcomp}, and introducing $\cp$ as in \eref{rel1}. From \eref{dtortresc} $r(\torc)$ is to be determined, or else first find $\torc(r)$ and then invert. When integrating \eref{dtortresc} for the compactified tortoise coordinate, the value of $\torc$ chosen for $r=1$ corresponding to \scrip does not make a difference in the Penrose diagrams. Here the integration choice $\torc|_{\scri^+}=r|_{\scri^+}=\rscri=1$ has been made, as well as imposing that the compactified coordinate takes the same value ($r_{hor}$) at the horizon for both branches of the compactified tortoise radial coordinate. The value of the latter at the trumpet is denoted as ${\torc}\trum$. 
The derivative of the height function in terms of the compactified tortoise coordinate is (imposing \eref{dtortresc} and \eref{hcinteg} in the second equality)
\begin{equation}\label{hptortresc}
h'(\torc) \equiv \partial_{\torc}h(\torc) = \frac{dr}{d\torc}\partial_{r}h(r) = -\frac{g_{tr}}{\Omega^2\cp} = -\frac{\gamma_{rr}\beta^r}{\chi\,\Omega^2\cp}. 
\end{equation}
The integration is performed as 
\begin{subequations}\label{integctort}
\begin{eqnarray}
&\pr>2M: \ h_{out} = -\int_{-\infty}^{1}\left[\frac{\gamma_{rr}\beta^r}{\chi\,\Omega^2\cp}\right](r_{out})d\torc, \label{integctortout}\\
&\pr<2M: \ h_{in} = -\int_{-\infty}^{\torc{}\trum}\left[\frac{\gamma_{rr}\beta^r}{\chi\,\Omega^2\cp}\right](r_{in})d\torc, \label{integctortin}
\end{eqnarray}
\end{subequations}
where $h_{out}$, $h_{in}$, $r_{out}$ and $r_{in}$ are functions of $\torc$. In practice, the lower limit of integration is taken as $\sim-8$ (or even a larger value, depending on the quality of the numerical data) instead of $-\infty$. The results are shown in \fref{heightct}, rescaled by the conformal factor. 

\bibliographystyle{unsrt}
\bibliography{../../hypcomp}

\end{document}